\documentclass[preprint,amsmath,amssymb,aps,]{revtex4-1}
 \usepackage{graphicx}
 \usepackage{graphicx}
 \usepackage{epstopdf}
 \usepackage{dcolumn}

 \usepackage{amsmath}
 \usepackage{bm}
 \usepackage{amssymb}
 \usepackage{amsfonts}

 \usepackage[noautoscale,vcentermath]{youngtab}

\begin{document}


\title{The SUSY Yang-Mills plasma in a $T$-matrix approach}

\author{Gwendolyn \surname{Lacroix}} 
\email[E-mail: ]{gwendolyn.lacroix@umons.ac.be}
\author{Claude \surname{Semay}}
\email[E-mail: ]{claude.semay@umons.ac.be}
\affiliation{Service de Physique Nucl\'{e}aire et Subnucl\'eaire,
Universit\'{e} de Mons -- UMONS, Place du Parc 20, 7000 Mons, Belgium}

\author{Fabien \surname{Buisseret}}
\email[E-mail: ]{fabien.buisseret@umons.ac.be}
\affiliation{Service de Physique Nucl\'{e}aire et Subnucl\'{e}aire,
Universit\'{e} de Mons -- UMONS,
Place du Parc 20, 7000 Mons, Belgium;\\ 
Haute \' Ecole Louvain en Hainaut (HELHa), Chauss\'ee de Binche 159, 7000 Mons, Belgium}

\begin{abstract}
The thermodynamic properties of ${\cal N}=1$ supersymmetric Yang-Mills theory with an arbitrary gauge group are investigated. In the confined range, we show that identifying the bound state spectrum with a Hagedorn one coming from non-critical closed superstring theory leads to a prediction for the value of the deconfining temperature $T_c$ that agrees with recent lattice data. The deconfined phase is studied by resorting to a $T$-matrix formulation of statistical mechanics in which the medium under study is seen as a gas of quasigluons and quasigluinos interacting nonperturbatively. Emphasis is put on the temperature range (1-5)~$T_c$, where the interaction are expected to be strong enough to generate bound states. Binary bound states of gluons and gluinos are indeed found to be bound up to 1.4 $T_c$ for any gauge group. The equation of state is then computed numerically for SU($N$) and $G_2$, and discussed in the case of an arbitrary gauge group. It is found to be nearly independent of the gauge group and very close to that of non-supersymmetric Yang-Mills when normalized to the Stefan-Boltzmann pressure and expressed as a function of $T/T_c$. 
\end{abstract}


\maketitle

\section{Introduction}

The phenomenology related to the QCD confinement/deconfinement phase transition has been the subject of intense investigation, both experimentally and theoretically (see \textit{e.g.} \cite{yagi} for a review of the topic). In comparison to QCD however, less information is known about the finite-temperature behavior of generic Yang-Mills (YM) theories, \textit{i.e.} with an other gauge group than SU(3) and matter in other representations than the fundamental one. Several results about a pure YM theory can be mentioned. First, there still exists a first-order deconfining phase transition with the gauge groups  G$_2$, SU($N>3$), SP(2) and E$_7$ \cite{G2,braun,Dumitru:2012fw}. Second, the equation of state (EoS) above the deconfining temperature $T_c$, appears to be nearly independent of the gauge group once normalized to the Stefan-Boltzmann pressure \cite{Dumitru:2012fw,Bruno,LSCB}. Below $T_c$, the EoS seems to be compatible with a Hagedorn-type density of state with the Hagedorn temperature $T_h$ playing the role of $T_c$ \cite{meyer09,Buisseret:2011fq}.

Among other extensions of YM theories, a particularly challenging case is the one with one flavor of massless Majorana fermions in the adjoint representation of the gauge group. Such a theory is supersymmetric and is the $\mathcal{N} = 1$ SUSY YM theory \cite{salam}, the adjoint quarks being called the gluinos. As in ordinary YM theory, the behavior of $\mathcal{N} = 1$ SUSY YM theory can be guessed from the $\beta$-function. It has been exactly computed from instanton calculus \cite{Novi83}, and reads $\beta(g)=-\frac{g^3}{16\pi^2}\frac{3N}{1-\frac{g^2 N}{8\pi^2}}$ for an SU($N$) gauge group. This form is compatible with asymptotic freedom, and it appears that $\mathcal{N} = 1$ SUSY YM is confining at $T=0$, as in the ordinary YM case. Several studies have been thus devoted to compute its spectrum with the gauge groups SU($N$) \cite{Farrar:1997fn,Campos:1999du,Farchioni:2004fy,Demmouche:2010sf,Bergner:2013nwa,Feo:2004mr}. Moreover, the theory is expected to exhibit a deconfining phase transition: Recent lattice results indicate that it is indeed the case, at least for SU(2) \cite{Bergner:2014saa}. This might be the case for an arbitrary gauge group too, according to \cite{Anber:2014lba}. At very high temperatures finally, the deconfined phase is expected to behave as a conformal gas of gluons and gluinos \cite{Amat88}.

A peculiarity of SU($N$) $\mathcal{N} = 1$ SUSY YM is that it is equivalent to one-flavor QCD at large $N$ provided that quarks are in the two-indices antisymmetric representation of the gauge group, which is isomorphic to the fundamental one at $N=3$. This duality is called orientifold duality and has attracted a lot of attention since the pioneering work \cite{Armo03}. 
  
The main purpose of the present work is to study the thermodynamic features of the deconfined phase of  $\mathcal{N} = 1$ SUSY YM theory by resorting to the formalism described in \cite{LSCB}. This formalism is based on a $T$-matrix approach that allows to model the bound states and scattering states of the system in an unified way \cite{cabr06}. From the $T$-matrix, the EoS can be computed by resorting to the Dashen, Ma and Bernstein's formulation of statistical mechanics~\cite{dashen}. Such a formulation is particularly well suited for systems whose microscopic constituents behave according to relativistic quantum mechanics: In our framework, the deconfined phase is seen as a strongly-interacting gas of gluons and gluinos propagating in the plasma. Note that the approach described in \cite{LSCB} has already proven to reproduce accurately the current lattice data concerning the EoS of ordinary YM theory for the gauge groups SU($N$).

The paper is organized as follows. Comments about the confined phase of $\mathcal{N} = 1$ SUSY YM and its critical temperature are first given in Sec. \ref{secconf}. The study of the deconfined phase will be made by resorting to the $T$-matrix formulation of statistical mechanics, which is presented in Sec.~\ref{Tmatsec}. Note that more detailed explanations about this approach can be found in \cite{LSCB}. The results obtained in the strongly-coupled stage of the deconfined phase, that is for temperatures just above $T_c$, are presented and discussed in Sec.~\ref{strcsec}. After some comments on the high temperature regime in Sec.~\ref{highT}, we show in Sec.~\ref{qcdasp} that our approach is compatible with the aforementioned orientifold equivalence. Our results are finally summarized in Sec.~\ref{conclu}, while technical details are given in the Appendices. 

\section{The confined phase}\label{secconf}

 To describe the high-lying hadronic spectrum, Hagedorn~\cite{hage65} proposed a model in which the number of hadrons with mass $m$ is found to increase as $\rho(m)\propto m^a \, {\rm e}^{m/T_h}$ ($a$ is real): the so-called Hagedorn spectrum. Thermodynamical quantities are then undefined for $T>T_h$. It is then tempting to guess that $T_h$ is the deconfining temperature, $T_c$ above which the new degrees of freedom are deconfined quarks and gluons rather than hadrons. 
 
Besides Hagedorn's original work, it is known that the degeneracy of string states versus their mass is also that of a Hagedorn spectrum, see \textit{e.g.} \cite{zwie}. This result has implications in YM theory: Glueballs are indeed frequently modelled by closed strings since the celebrated Isgur and Paton's flux tube model \cite{isgur}, inspired from the Hamiltonian formulation of lattice QCD at strong coupling. According to this closed-string picture of glueballs, it has been shown that the EoS of pure SU(3) YM theory computed on the lattice is compatible with a glueball gas model in which the high-lying spectrum is modelled by the one of a closed bosonic string~\cite{meyer09}. 

In the non-SUSY YM case, identifying the critical temperature to the Hagedorn temperature of a bosonic closed string theory in $(3+1)$-dimensions agrees well with currently known lattice data \cite{meyer09,Buisseret:2011fq,hage2}: $T_c(\textrm{non-SUSY)}/\sqrt\sigma=\sqrt{3/(2\pi)}\approx 0.7$. Correspondingly, in ${\cal N}=1$ SUSY YM, we conjecture that the Hagedorn temperature should be that of a non-critical (\textit{i.e.} well-defined in a 4-dimensional spacetime) closed superstring theory. Such a theory has been studied in particular in \cite{susystr}, where the usual Hagedorn temperature is recovered for the bosonic case and where the ratio 
\begin{equation}\label{TC}
\frac{T_c(\textrm{SUSY})}{T_c(\textrm{non-SUSY})}=\sqrt{\frac{2}{3}}\approx 0.8
\end{equation}
is found for the superstring. Interestingly the same value has been recently found in a SU(2) lattice simulation of ${\cal N}=1$ SUSY YM thermodynamics \cite{Bergner:2014saa}. Equation~(\ref{TC}) thus provides an explanation to this value, finally leading us to the result 
\begin{equation}\label{Tcdef}
T_c(\textrm{SUSY})/\sqrt\sigma=1/\sqrt{\pi}\approx 0.6 , 
\end{equation} 
that will be used throughout this paper. 

\section{Deconfined phase and the $T$-matrix formalism}\label{Tmatsec}

\subsection{Generalities}\label{gen}

The results of Ma, Dashen and Bernstein \cite{dashen} establishing the grand potential of an interacting relativistic particle gas, $\Omega$, expressed as an energy density, are summarized by the following virial expansion (in units where $\hbar=c=k_B=1$).
\begin{equation}\label{pot0}
\Omega=\Omega_0+\sum_\nu\left[\Omega_\nu-\frac{{\rm e}^{\beta\bm \mu\cdot\bm  N}}{2\pi^2\beta^2}\int^\infty_{M_\nu} \frac{dE}{4\pi i}\, E^2\,  K_2(\beta E)\, \left. {\rm Tr}_\nu \left({\cal S}S^{-1}\overleftrightarrow{\partial_E}S \right)\right|_c\right]\text{.}
\end{equation}  
In this equation, the first term, $\Omega_0$, is the grand potential of the free relativistic particles. The second term accounts for interactions in the plasma and is a sum running on all the species, the number of particles included, and the quantum numbers necessary to fix a channel. The set of all these channels is generically denoted $\nu$. The vectors $\bm \mu=(\mu_1, \mu_2,\dots)$ and $\bm N=(N_1,N_2,\dots)$ contain the chemical potentials and the particle number of each species taking part in a given scattering channel. 

In \eqref{pot0}, we can notice that the contribution of the bound states and scattering states are decoupled. Below the threshold $M_\nu$ (the threshold is a summation on the mass of all the particles included in a given channel $\nu$), bound states appearing in the $S$-matrix spectrum are added as free additional species: $\Omega_\nu$ is the grand canonical potential describing an ideal relativistic gas of the $\nu$-channel bound states. Above $M_\nu$, the scattering contribution is expressed as an integration depending on a trace, taken in the center of mass frame of the particles in the channel $\nu$, and is a function of the $S$-matrix, $S$, of the system. $S$ is in particular a function of the total energy $E$. The symmetrizer ${\cal S}$ enforces the Pauli principle when a channel involving identical particles is considered, and the subscript $c$ means that only the connected scattering diagrams are taken into account. $K_2(x)$ is the modified Bessel function of the second kind and $\beta = 1/T$ where $T$ is the temperature. The notation $A\overleftrightarrow{\partial_x} B$ denotes $A(\partial_xB)-(\partial_xA)B$.

By definition, $S$ is linked to the off-shell $T$-matrix,  ${\cal T}$, by the relation
\begin{equation}
S=1-2\pi i\, \delta(E-H_0)\, {\cal T}\text{,}
\end{equation} where $H_0$ is the free Hamiltonian of the system. A way to obtain ${\cal T}$ in two-body channels is to solve the Lippmann-Schwinger equation, schematically given by
\begin{equation}\label{ls}
{\cal T}=V+ V\, G_0\, {\cal T} \text{,}
\end{equation}
with $G_0$ the free propagator and $V$ the interaction potential. At this point we have to mention that, since we will focus on two-body channels only, the Lippmann-Schwinger equation is the relevant one to use. In three-body channels for example, Faddeev equations should be used instead in order to eliminate the spurious solution of the Lippmann-Schwinger equation \cite{joachain}.

Once Eq.~(\ref{pot0}) is computed, all thermodynamic observables can be derived. For example, the pressure is simply given by 
\begin{equation}
p=-\Omega .
\end{equation}

The sum $\sum_\nu$ appearing in (\ref{pot0}) explicitly reads $\sum_{n}\sum_{I}\sum_{J^{PC}}\sum_{{\cal C}}$, where $n \ge 2$ is the number of particles involved in the interaction process, $I$ is a possible isospin channel, ${\cal C}$ is the color channel, and $J^{PC}$ is the spin/helicity channel (the labels $C$ or $P$ must be dropped off if the charge conjugation or the parity are not defined). As in \cite{LSCB}, we assume that the dominant scattering processes are the two-body ones, and thus we only consider $n=2$.

The normalized trace anomaly can also be computed by the following formula
\begin{equation} \label{ta}
\frac{\Delta}{p_{SB}} =  -\beta \left( \partial_\beta \frac{p}{p_{SB}} \right)_{\beta\mu}.
\end{equation}
Within this paper, we will give some results about the trace anomaly, as an indication. It is indeed mentioned in \cite{LSCB} that some improvements must be done in order to obtain a reliable estimation of this quantity. 

\subsection{Quasiparticle properties} \label{properties}

A key ingredient of the present approach is the 2-body potential $V$, encoding the interactions between the particles in the plasma. $V$ is chosen as in \cite{LSCB}: It is extracted from the free energy $F_1$ between a $q\bar{q}$ pair at finite temperature, computed in quenched SU(3) lQCD \cite{lPot}. Note that unquenched results are available, but only small differences appear with respect to quenched calculations \cite{Kacz05}. 

For the needs of our approach, this free energy has been fitted with a Cornell potential screened thanks to the Debye-H\"uckel theory \cite{Satz90} (see Appendix~B in \cite{LSCB}). However as in  \cite{LSCB}, we prefer to use the internal energy $U_1$ as the interaction potential. This choice is still a matter of debate. Nevertheless, it has given correct results in the ordinary YM case, as shown in  \cite{LSCB}. 

The Casimir scaling seems very well respected between two static color sources in the $T=0$ sector \cite{bali}. Computations in the $T>T_c$ sector show a situation which is slightly different: The Casimir scaling seems partly violated (at most 20\%) for short distances and temperatures near $T_c$ \cite{gupta}. In this work, as in \cite{LSCB}, we assume that the Casimir scaling is verified. We use the lattice $q\bar{q}$-potential, as proposed in Sec.~II in \cite{LSCB}, and we rescale the color dependence to obtain the following form
\begin{equation}
V(r,T) = \displaystyle\frac{\kappa_{{\mathcal{C};ij}}}{\kappa_{ \bullet ; q\bar{q}}} \left[U_1(r,T) - U_1(\infty , T) \right], 
\label{Used_pot}
\end{equation}
\noindent where 
\begin{equation}
\kappa_{{\mathcal{C};ij}} = \displaystyle\frac{C_2^{\mathcal{C}} - C_2^{R_i} - C_2^{R_j}}{2 C_{2}^{\text{adj}}},
\label{color_scaling}
\end{equation}
and where $C_2^R$ is the quadratic Casimir of the representation $R$. $\mathcal{C}$, adj, $\bullet$, $R_i$ and $R_j$  stand respectively for the pair, adjoint, singlet, $i$- and $j$-particle representation. For instance, 
\begin{equation}
C_{2}^{\text{adj}} = N \text{, } C_2^{\bullet} = 0 \text{, } C_2^{q} = C_2^{\bar{q}} = \displaystyle\frac{N^2 - 1}{2N} ,
\end{equation}
for the SU($N$) gauge group, where $q$ denotes the fundamental representation. The values taken by (\ref{color_scaling}) for the various color channels considered in this paper are given in Tables \ref{tab:c0}-\ref{tab:c3} in Appendix~\ref{colorchan}. Let us note the presence of $\kappa_{{\bullet; q\bar{q}}}=-4/9$, since $U_1(r,T)$ is fitted on a singlet $q\bar{q}$ potential for SU(3) \cite{LSCB}.

The color dependence respecting the Casimir scaling is the simplest form possible for two color sources. It has the same form as the one for the one-gluon exchange process, but our interaction contains other processes since it stems from a lQCD computation. 
In this work, all hyperfine interactions are neglected. We can expect that they are non-dominant with respect to the spin-independent contributions, since these processes are assumed to depend on the inverse square of the effective mass. With this hypothesis, we also miss the diagonal annihilation contributions. Let us note that the annihilation mechanism, which does not respect the Casimir scaling, is a contact interaction and is then vanishing for all non-S states. 

In \eqref{Used_pot}, the long-distance behavior of the lattice potential $U_1(\infty, T)$, is subtracted since this term is assimilated, as suggested in \cite{Mocs06}, as a thermal mass contribution for the quasiparticles (this also ensures the convergence of the scattering equation and the possibility to perform the Fourier transform). Indeed, when the quasiparticles are infinitely separated, the only remaining potential energy can be seen as a manifestation of the in-medium self-energy effects, $U_1(\infty, T) = 2m_q(T)$. We thus encode these effects as a mass shift $\delta(T)$ to the ``bare" quasiparticle mass $m_0$, by following the arguments exposed in \cite{LSCB}:
\begin{equation}
m(T)^2 = m_{0}^2 + \delta(T)^2 ,
\label{mg}
\end{equation}
 where $m_0$ is independent of $T$. In order to get the thermal mass for any particles, we extract the first-order color dependence in agreement with the hard-thermal-loop (HTL) leading-order behavior \cite{Blai99}: 
\begin{equation}
\delta(T) = \sqrt{\frac{C_2^R}{C_2^{\text{adj}}}} \Delta (T) ,
\end{equation}
where the quantity $\Delta (T)$ is assumed to be color-independent. As $U_1(r,T)$ is fitted on a singlet $q\bar{q}$ potential for SU(3), we have here
\begin{equation}
\displaystyle\frac{U_1(\infty,T)}{2} = m_q(T) = \sqrt{\frac{C_2^q}{C_2^{\text{adj}}}} \Delta (T) = \frac{2}{3} \Delta (T).
\end{equation}
In particular, $\delta(T) = \Delta (T)$ for the gluon and the gluino since they both belong to the adjoint representation of the gauge group. It comes naturally that $\delta(T)$ for these particles is gauge-group independent in our model. Looking at the behavior of $m(T)$ (see Sec.~V in \cite{LSCB}), it can be seen that the particles become more and more light with increasing values of $T$. So, they are less and less influenced by the short range part of the interaction (where the Casimir scaling is less satisfied). Moreover, in this case, the Casimir scaling is more and more satisfied. So we can expect that the fundamental hypothesis of our model is better verified at high values of $T$. 

\subsection{Solving Lippman-Schwinger equations}\label{LSeq}

Having established the quasiparticle properties, we have now all the ingredients to solve the Lippman-Schwinger equation leading to the on-shell $T$-matrix. It can be computed from (\ref{ls}) as follows in \cite{LSCB}: 
\begin{eqnarray}\label{tosolve}
{\cal T}_\nu(E; q,q') &=& V_\nu(q,q') + \frac{1}{8\pi^3} \int_0^\infty dk\, k^2\, V_\nu(q,k)  \\
&& \times \,G_0(E;k)\, {\cal T}_\nu(E;k,q') \left[1 \pm f_{p_1}(\epsilon_1)\right] \left[1 \pm f_{p_2}(\epsilon_2)\right], \nonumber
\end{eqnarray}
where $E$ is the energy in the center-of-mass frame, $\epsilon_i$ the asymptotic energy of the particle $i$, and where the free two-body propagator is given by~(\ref{G0}). The symbol $\nu$ stands for the particular channel considered. 

For ordinary $\left|^{2S+1}L_J\right\rangle$ states, $V_\nu(q,q')$ is given by the Fourier transform of the interaction  
\begin{equation}
V_L(q,q') = 2\pi \displaystyle\int_{-1}^{+1} dx P_L(x) V(q,q',x),
\end{equation}
where $P_L$ is the Legendre polynomial of order $L$, and $x = \cos \theta_{q,q'}$ with $\theta_{q,q'}$ the angle between the momenta $\vec q$ and $\vec q\,'$. The spin $S$ is not indicated since our interaction is spin-independent. Our potential has a spherical symmetry. We have then
\begin{equation}\label{Vqq}
V(q,q',\theta_{q,q'}) = 4 \pi \displaystyle\int_0^\infty dr \, r V(r) \displaystyle\frac{\sin(Q\,r)}{Q} ,
\end{equation}
where $Q = \sqrt{q^2 + q'^2 - 2 q q' \cos \theta_{q,q'}}$. 

When at least one particle is transverse, we have to use the helicity formalism \cite{jaco}. It is very convenient to use, since a particular helicity state $\left|J^P\right\rangle$ can be written as (see Appendix~\ref{helicity})
\begin{equation}\label{helexpspin}
\left|J^P\right\rangle = \sum_{L,S} C_{L,S} \left|^{2S+1}L_J\right\rangle. 
\end{equation}
Then, it can be shown that 
\begin{equation}
V_{J^P}(q,q') = \sum_{L,S} C_{L,S}^2 V_L(q,q'),
\end{equation}
since our interaction is spin-independent. 

Contrary to the $T$-matrix equation used in \cite{LSCB}, we have included the in-medium effects, namely the Bose-enhancement and the Pauli-blocking. According to \cite{Prat94}, these in-medium effects change the cross-section as follows, 
\begin{equation}
\sigma^{med} = \sigma^{vac} (1 \pm f_{p_1}) (1  \pm f_{p_2}) \text{,}
\label{med_effects}
\end{equation}
where $\sigma^{vac}$ and $\sigma^{med}$ are respectively the cross-section in the vacuum and in the medium, and where $f_p$ is the distribution function of the $p$-species. If the species is a boson, 
\begin{equation}
f_p (\epsilon) = \frac{1}{e^{\beta(\epsilon - \mu)} - 1} ,
\end{equation}
while if the species is a fermion, 
\begin{equation}
f_p (\epsilon)= \frac{1}{e^{\beta(\epsilon - \mu)} + 1} ,
\end{equation}
where $\mu$ is a possible chemical potential. 
The sign choice in (\ref{med_effects}) also depends on the nature of the particles: $+$ for bosons and $-$ for fermions. Let us note that these in-medium effects have a very small influence on the EoS in our computations. Therefore, the results obtained in \cite{LSCB} remain valid.

The Haftel-Tabakin algorithm is used to solve (\ref{tosolve}) \cite{haftel}. The
momentum integral is discretized within an appropriate quadrature, thus turning
the integral equation in a matrix equation, namely, $\sum {\cal F}_{ik} {\cal
T}_{kj} = V_{ij}$, where, schematically \cite{cabr06}, 
\begin{equation}
{\cal F}=1-wVG (1 \pm f_{p_1}) (1  \pm f_{p_2})
\end{equation}
and $w$ denotes the integration weight. The solution follows trivially by matrix inversion. It can be shown that the determinant of the transition function $\cal F$ (referred to
as the Fredholm determinant) vanishes at the bound state energies, which provides
a numerical criterion for solving the bound state problem. This strategy has
already been successfully used to compute $T$-matrices in \cite{LSCB,cabr06}.
Once ${\cal T} (E;q,q')$ is known, the on-shell $T$-matrix is readily obtained
as ${\cal T}(E;q(E),q(E))$, with $q(E)$ given by (\ref{diff_mass}). 

The Haftel-Tabakin algorithm is a reliable procedure to solve the $T$-matrix problem \cite{cabr06,haftel}. Recently, invoking the Levinson's theorem, it has been shown that keeping only the bound-state spectrum in thermodynamical computations leads to strong violations of unitarity \cite{werg03}. In our calculations, we take into account the continuum as well the bound states, in order to keep unitarity. In a previous work \cite{LSCB}, we have checked that the number of nodes that ${\rm Re} {\cal T}_{\nu}(E)$ presents in the scattering region changes in parallel with the number of bound states. These behaviors are the reflections of Levinson's theorem in our calculation. 

\subsection{Computing the equation of state}\label{ComputEoS}

The first term in (\ref{pot0}) is the free relativistic gas. It is given by 
\begin{equation}
\Omega_0 = \sum_{\textrm{species}}(2 I+1) \times {\rm dim}\ J \times {\rm dim}\ {\cal C} \times \omega_\pm(m).
\end{equation}
The sum runs on each species inside the plasma. $I$ is the isospin of the particle, ${\rm dim}\ J$ is the dimension of the spin/helicity representation, and ${\rm dim}\ {\cal C}$ is the dimension of the color representation. Moreover,
\begin{equation}
\omega_+(m) = \frac{1}{2\pi^2\beta}\int^\infty_0dk\, k^2 \ln\left(1-{\rm
e}^{-\beta\sqrt{k^2+m^2}}\right) 
\end{equation}
is the grand potential per degree of freedom associated to a bosonic species with mass $m$, while
\begin{equation}
\omega_-(m) = -\frac{1}{2\pi^2\beta}\int^\infty_0dk\, k^2 \ln\left(1+{\rm
e}^{-\beta\sqrt{k^2+m^2}}\right) 
\end{equation}
is the grand potential per degree of freedom associated to a fermionic species with mass $m$. For later convenience, the thermodynamic quantities will be normalized to the Stefan-Boltzmann pressure, which is defined as 
\begin{equation}
p_{SB} = - \displaystyle\lim_{m \rightarrow 0} \Omega_0\text{.}
\label{SB}
\end{equation}

Attractive interactions in a particular channel $\nu$ can lead to the formation of bound states with masses $\mu_\nu < m_1+m_2$. As mentioned above, the energy of these states can be computed by looking at the zeros of the determinant of the transition function $\cal F$. They contribute also to the grand potential as new species via the formula 
\begin{equation}
\sum_\nu \Omega_\nu = \sum_{\textrm{attractive channels}}(2 I_\nu+1) \times (2 J_\nu+1) \times {\rm dim}\ {\cal C}_\nu \times \omega_\pm(\mu_\nu).
\end{equation}
In this sum, $\omega_\pm(\mu_\nu)=0$ if no bound state exists in the channel $\nu$ at a given value of $T$. The full form of the scattering part of (\ref{pot0}) is given in Appendix~\ref{diff} and is not recalled here for the sake of clarity.

\begin{figure}[h!]
\begin{center}
\includegraphics*[width=0.7\textwidth]{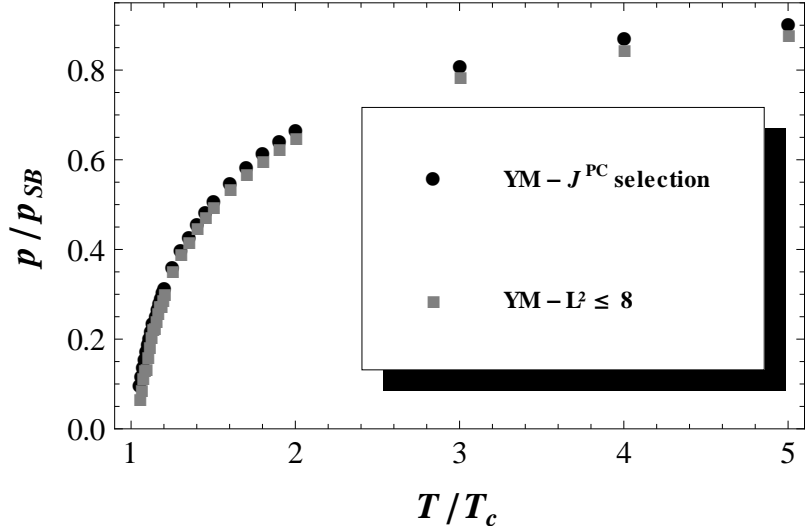}
\end{center}
\caption{Normalized pressure of a pure YM plasma for two different sets of $gg$ channels: all $J^{PC}$ channels considered in this paper with our criteria given above (circle); all channels such that $\left\langle \bm L^2\right\rangle \le 8$ (square). Parameters are given in Sec. \ref{secparam}, except for $T_c=300$~MeV \cite{LSCB}. }
\label{selection}
\end{figure}

For obvious numerical reasons, all the possible channels contributing to $\Omega$ can not be included into the sum~(\ref{pot0}). So we need reliable criteria to select the most significant channels. When the total spin $J$ increases, the average value $\left\langle \bm L^2\right\rangle$ increases also. This is obvious for ordinary spin states, and this is shown in Appendix~\ref{helicity} for helicity states. For a bound state, this means the increase of the mass, and then a reduced contribution to the grand potential. Moreover, in a naive nonrelativistic picture, the strength of the orbital barrier increases with $\left\langle \bm L^2\right\rangle$ in a scattering process, which reduces the value of the corresponding $T$-matrix~(\ref{tosolve}) (see Appendix~\ref{cross}). So we decided to restrict the sum~(\ref{pot0}) to channels with the lowest values of $\left\langle \bm L^2\right\rangle$. More precisely, a mean cross section $\bar \sigma_{J^P}$ is computed for each channel (see Appendix~\ref{cross}). Are only retained, the channels for which the value of $\bar \sigma_{J^P}$ is at least 25\% of the value $\bar \sigma_{J^P}$ for the channel with the lowest value of $\left\langle \bm L^2\right\rangle$. In this case, nearly all states with $\left\langle \bm L^2\right\rangle \le 5$ are retained in the sum. We have checked that the value of 25\% is a good compromise between accuracy and computational effort. For instance, the inclusion of all $gg$ states up to $\left\langle \bm L^2\right\rangle=8$ brings very weak modifications to the pressure, as it can be seen on Fig.~\ref{selection}.  

\subsection{$T = 0$ bound states} \label{T0BS}

With the formalism described above, it is possible to compute bound states within the plasma (see Sec.~\ref{LSeq}). Nevertheless, it is also interesting to compute bound states at $T=0$, in order to check the validity of our model. In quenched SU(3) lattice QCD, the potential between a static quark-antiquark pair at zero temperature is compatible with the funnel form \cite{bali}
\begin{equation}
V_f(r) = \sigma r -\frac{4}{3}\frac{\alpha}{r}, 
\end{equation}
where $\alpha \sim 0.4$  and $\sigma \sim 0.2$~GeV$^2$ (standard values for the running coupling constant $\alpha$ and the string tension $\sigma$ at $T=0$). Again, we neglect the contributions of annihilation processes. Since the Fourier transform of $V_f(r)$ is not defined (because of a nonzero asymptotic value), a string-breaking value, $V_{sb}$, is introduced to make it convergent. $V_{sb}$ is thus seen as the energy above which a light quark-antiquark pair can be created from the vacuum and breaks the QCD string. This scale is then subtracted and the potential effectively taken into account is $V_f(r)-V_{sb}$, while $V_{sb}/2$ is interpreted as an effective quark mass using the same arguments as those detailed in Sec.~\ref{properties} \cite{cabr06}. 

According to the color scaling (\ref{color_scaling}), the potential describing the interactions between two color sources (with representations $R$ and $\bar R$) at zero temperature, and thus only in a singlet gauge representation ($C_2^{\bullet} = 0$) since confinement is present, is  
\begin{equation}\label{T0pot}
V_0(r) = \frac{9}{4}\left( C_2^{R} + C_2^{\bar R} \right) V_f(r) - V_{sb}^R .
\end{equation}
Let us recall that the factor $9/4$ appears since the potential $V_f$ is fitted on a singlet $q\bar{q}$ pair for SU(3). In this case, $V_{sb}^R$ should rather be interpreted as the energy scale necessary to form two sources of color compatible with the existence of the two new color singlet pairs of particles created by the string breaking. For instance, it is expected that two gluelumps can be formed for the breaking of a string between two gluons (a gluelump is a gluon bound in the color field of a static adjoint source). If $m_0$ is the bare mass of the particle, the $T=0$ mass $m(0)$ used to compute the bound state is then
\begin{equation}
m(0)^2 = m_0^2 + \left(\displaystyle\frac{V_{sb}^R}{2}\right)^ 2,
\end{equation}
keeping the same structure as in (\ref{mg}). This procedure is the same as the one used in \cite{LSCB}.

\section{The strongly-coupled phase}\label{strcsec}
\subsection{Parameters and assumptions}\label{secparam}
Now, we will particularize the general formalism presented in the previous section for an $\mathcal{N} = 1$ SUSY YM theory. In such a theory there are two species of quasiparticles: the gluons ($g$) and their supersymmetric partners, the gluinos ($\tilde{g}$). Both particles have a vanishing isospin. Note that the nonsupersymmetric case was treated in \cite{LSCB}. The gluons are transverse spin-1 bosons in the adjoint representation of the gauge group and thus, since the gluinos are their supersymmetric partners, they are Majorana fermions of spin 1/2 belonging to the same representation. The two-body channels to be considered are $gg$, $\tilde{g}\tilde{g}$ and $g\tilde{g}$. The lowest corresponding spin/helicity states are given in Appendix \ref{helicity}, and the possible color channels can be found in Appendix~\ref{colorchan}. 

The possible interactions are then diagonal ones in $gg$, $\tilde{g}\tilde{g}$ and $g\tilde{g}$ pairs, as well as transition processes between $gg$ and $\tilde{g}\tilde{g}$ pairs. Let us examine the different possibilities: 
\begin{itemize} 
\item As explained above, the interaction (\ref{Used_pot}) between two gluons or two gluinos follows strictly the Casimir scaling and neglects all hyperfine corrections, annihilation ones included. 
\item Although this interaction is expected to take into account complicated exchanges (since it stems from a lQCD calculation), it is interesting to look at the simplest possible Feynman diagrams between two particles. Two gluons or two gluinos can exchange a gluon, but the basic gluon-gluino interaction is a gluino exchange. So the choice (\ref{color_scaling}) for the color factor is questionable for this particular interaction. To correct this point is beyond the scope of this work, but it is worth mentioning that the contributions of the $\tilde g g$ interactions is expected to be very weak in our model (see Sec.~\ref{highT}). 
\item Processes transforming a $gg$ pair into a $\tilde{g}\tilde{g}$ pair can exist, but we have checked that mechanisms of order 1 are naturally suppressed since there is no overlap between $gg$ and $\tilde{g}\tilde{g}$ states (see Sec.~\ref{tildegtildeg}). As we neglect second order processes, as hyperfine interactions, we do not take into account transition between $gg$ and $\tilde{g}\tilde{g}$ pairs.
\end{itemize}

Unbroken supersymmetry is assumed in this paper, so the parameters can be fixed assuming that $m_g=m_{\tilde{g}}$ (so, $m_{0g}=m_{0\tilde{g}}=m_0$) and the interaction potential is the one used in \cite{LSCB}, independent of the interacting species, gluon or gluino. 

The string tension $\sqrt{\sigma}$ can be taken as the fundamental scale of energy, so that the meaningful parameters are $m_0/\sqrt{\sigma}$ and $T_c/\sqrt{\sigma}$. This last ratio is defined by Eq. (\ref{Tcdef}) for any gauge group. The gluon bare mass value $m_0/\sqrt \sigma=1.67$ is found by matching our $\cal T$-matrix results and the lattice ones in the bound state sector at $T=0$ of the non-SUSY YM case with gauge group SU(3), see \cite{LSCB}. Moreover, it is an acceptable value for the zero-momentum limit of the gluon propagator at $T = 0$ (see e.g. \cite{Oliv11,Corn82,Bino09}). The ratio $m_0/\sqrt \sigma$ is kept for any gauge group since all the dependence of the masses on the gauge group is assumed to come from the definition (\ref{mg}). This assumption is coherent with the lattice study \cite{Maasglu}, where the gluon propagator in Landau gauge has been shown to be nearly independent of the gauge group once normalized to the string tension. 

Finally, we take $\alpha=0.4$ at $T=0$ and $\alpha=0.141$ for $T > T_c$. With $\sigma=0.176$~Gev$^2$, we have $T_c=0.25$~GeV and $m_0=0.7$~GeV.

\subsection{Bound states at zero temperature}

The zero temperature spectrum of the theory can be computed by solving (\ref{tosolve}) with potential~(\ref{T0pot}). Although our main goal is the study of the deconfined phase of the theory, computing the zero temperature spectrum has an interest in view of comparing with currently known results in this field. Our results are given in Table~\ref{T0BStab} for the lightest two-body bound states we find in the singlet channel. Note that, in our formalism, these masses are independent of the number of colors, as already shown in \cite{LSCB}. 

\begin{table}[ht]
\caption{Masses (in GeV) of the lowest-lying bound states at $T=0$ with the gauge group SU($N$) in ${\cal N}=1$ SUSY YM. $J^{PC}$ is indicated for $gg$ and $\tilde g\tilde g$ states. The notation defined in Appendix~\ref{helicity} is used to further characterize the $\tilde g g$ states, although parity is meaningless in this case \cite{zuk83}. The typical value $\sigma=0.176$ GeV$^2$ is chosen.}
\begin{center}
\begin{tabular}{ccc}
\hline\hline
Content  & State & Mass \\
\hline
$\tilde g\tilde g$ & $0^{-+}$ & 1.58 \\
 $\tilde g g$ & $\frac{1}{2}^-$   & 1.96 \\
$gg$ & $0^{++}$ &  1.96  \\
 \hline 
 $\tilde g\tilde g$  & $ 0^{++}$ & 2.26 \\
   $\tilde g g$ & $\frac{1}{2}^+$  & 2.26 \\
 $gg$ &$0^{-+}$ &  2.26 \\
 \hline
  $\tilde g g$  & $\frac{3}{2}^-$ & 2.13 \\
$gg$&$2^{++}$ & 2.21  \\
 $\tilde g\tilde g$  & $ \{1,2\}^{++}$ & 2.26 \\
  $\tilde g g$   & $\frac{3}{2}^+$ & 2.31 \\
\hline\hline
\end{tabular}
\end{center}
\label{T0BStab}
\end{table}

It is known from effective Lagrangians that supersymmetry is not expected to be broken at the level of the bound state spectrum. Hence the lightest states should form two supermultiplets \cite{Farrar:1997fn}. Since hyperfine corrections (spin-spin, spin-orbit, etc.) are neglected in the present work, we expect these supermultiplets to be observable but only approximately degenerated. 

The first supermultiplet can be seen in the first three lines of Table \ref{T0BStab}. As expected from \cite{Farrar:1997fn}, it contains the pseudoscalar $\tilde g\tilde g$  state, also known as the $a-\eta'$ (the adjoint $\eta'$), a spin $1/2$ state and the scalar glueball. The last two states are degenerate, but not the $a-\eta'$ which is lighter. This is mainly a consequence of the approach used here: The $a-\eta'$ being a pure $S$-wave, it is maximally sensitive to the attractive Coulomb interaction and its mass is logically smaller than the other state ones. The degeneracy could be recovered by including for example spin-orbit terms that would decrease the mass of the $1/2^-$ and $0^{++}$ states only, but this is out of the scope of the present calculations.  

The second supermultiplet is shown in the lines four to six of Table \ref{T0BStab}: The masses are equal as expected from supersymmetry. We have also listed other states with higher spins: They are absent of low-energy effective actions but straightforwardly computed within our formalism. 
 
Our spectrum is actually very similar to what is observed in lattice studies, see for example \cite{Farchioni:2004fy,Demmouche:2010sf,Bergner:2013nwa}. In these studies also, the three lightest states are not exactly degenerate although very close up to the error bars, and the $a-\eta'$ is the lightest state. It is an indication that exact supersymmetry is still not reached on the lattice (or SUSY is spontaneously broken), as in our model. Smaller lattice sizes would be needed in order to draw definitive conclusions on the structure of the spectrum \cite{Bergner:2013nwa}, so the agreement between our model and the lattice data should be, in our opinion, restricted to qualitative considerations. 

We finally mention the work \cite{Feo:2004mr}, in which information on the ${\cal N}=1$ SUSY YM spectrum is obtained by resorting to the orientifold duality between the theory under study and QCD with one quark flavor. In such an approach, the $a-\eta'$ could be the lightest state of the theory without being degenerated with the other states of the aforementioned supermultiplet. So the zero temperature mass spectrum is still an open problem in our opinion.

\subsection{Bound states above $T_c$}

The existence or not of bound states in the deconfined phase is not forbidden in principle, especially around $T_c$ where interactions are strong enough to bind two or more particles. Within our formalism, the channels in which bound states are favored at most should contain a $S$-wave component to avoid the centrifugal barrier and should have a symmetry that allows the state to be in a color singlet, the color channel in which the interactions are maximally attractive. 

In the $gg$ case there are two such states: The $0^{++}$ and $2^{++}$ color singlets, corresponding to the scalar and tensor glueballs respectively. We have observed in \cite{LSCB} that both the scalar and tensor glueball masses at $1.05$ $T_c$ were compatible with the zero-temperature ones. Moreover the scalar glueball exists as a bound state up to 1.25 $T_c$ while the tensor one is bound up to 1.15 $T_c$. These results are independent of the number of colors and are still valid in the supersymmetric extension that we study in the present paper, since the influence of the Bose-enhancement appears very small.

The $\tilde g \tilde g$ case appears to be promising to find bound states because the $0^{-+}$ channel is a pure $S$-wave that can exist in the singlet, the $a-\eta'$. As in the glueball case, the $a-\eta'$ has a mass close to its zero temperature value in 1.05 $T_c$. It survives as a bound state up to 1.40 $T_c$; it is actually the most strongly bound state within our framework. The $a-\eta'$ can also exist in the symmetric adjoint color channel when the number of colors is larger than 2. This channel being less attractive, the dissociation temperature is lower and is equal to 1.20 $T_c$.  Complete results are listed in Table \ref{tab1}. Note that $\{0,1,2\}^{++}$ $\tilde g \tilde g$ stats also survive at 1.05 $T_c$, but they disappear immediately above this temperature. That is why they are not listed in Table \ref{tab1}.

In the $\tilde g g$ case, the ground state is formed in the $1/2^-$ channel. The comments that can be made are very similar to the two previous cases: The mass at 1.05 $T_c$ is nearly equal to that at $T=0$, and the dissociation temperature is equal to 1.30 $T_c$  in the color singlet and to 1.10 $T_c$ in the adjoint channel. 

\begin{table}[h]
\caption{Masses (in GeV) of some bound states above $T_c$. A line marks the temperature at which a bound state is not detected anymore. The typical value $\sigma=0.176$ GeV$^2$ is chosen.}
\begin{center}
\begin{tabular}{cccc}
\hline\hline
$T/T_c$ & $0^{-+}$ ($\tilde g\tilde g$) & $1/2$ ($\tilde g g$) & $0^{++}$ ($gg$) \\
\hline
1.05 & 1.30 & 1.90 & 1.90  \\ 
1.10 & 1.67 & 1.92 & 1.91 \\
1.15 & 1.72 & 1.87 & 1.86 \\ 
1.20 & 1.73 & 1.82 & 1.80 \\ 
1.25 & 1.71 & 1.78 & 1.77  \\
1.30 & 1.71 &  -   & - \\
1.35 & 1.70 &  \   & \  \\
1.40 & -    &  \   & \ \\
\hline\hline
\end{tabular}
\end{center}
\label{tab1} 
\end{table}

In summary, our computations show that two-body bound states can form in the gluino-gluon plasma for any species in the range (1-1.40) $T_c$. The picture developed in the pioneering work \cite{Shuryak:2004tx}, where the importance of binary bound states on the features of the quark-gluon plasma was stressed, seems thus to be valid in a supersymmetric extension of YM theory too. A more detailed look at Table \ref{tab1} shows that the behavior of a given bound state mass with the temperature is not systematic: It can either increase or decrease before dissociation. We have pointed out in \cite{LSCB} that the observed behavior in the glueball sector is in qualitative agreement with the lattice study \cite{Meng:2009hh}. As soon as gluinos are involved, there are, to our knowledge, no study to which our results can be compared. 

\subsection{Equation of state of ${\cal N}=1$ SUSY YM}\label{EoSsec}

\begin{figure}[h!]
\begin{center}
\includegraphics*[width=0.7\textwidth]{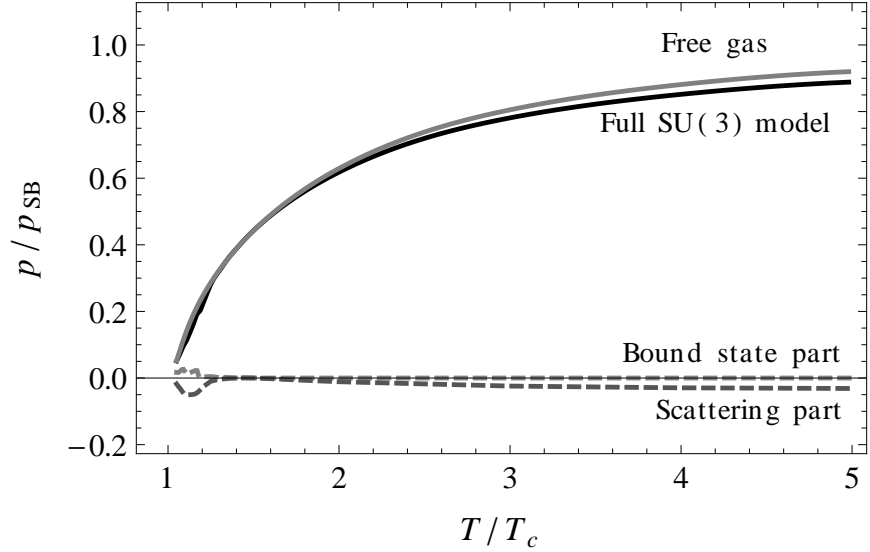}
\caption{Normalized pressure $p/p_{SB}$ versus temperature in units of $T_c$, computed for the gauge group SU(3). The full pressure is shown, together with the free gas, bound states and scattering contributions. From this figure to Fig.~\ref{anom_Susy}, results are given for for ${\cal N}=1$ SUSY YM system and parameters are given in Sec. \ref{secparam}.}
\label{Contri_Susy}
\end{center}
\end{figure}

We are now in position to compute the pressure of ${\cal N}=1$ SUSY YM normalized to the Stefan-Boltzman pressure, which reads $p_{SB}=\pi^2 T^4 (N^2-1) /24$ in the present case. We focus in this part on the temperature range (1-5) $T_c$, that should correspond to the strongly-coupled phase of the gluon-gluino plasma. The full SU(3) pressure is displayed in Fig.~\ref{Contri_Susy} and decomposed into the free gas, scattering and bound state parts. The free part is given by
\begin{equation}\label{Om0scr}
\Omega_0 = 2 (N^2-1)\left[ \omega_+(m_g) + \omega_-(m_{\tilde g}) \right],
\end{equation}
with $m_g = m_{\tilde g}$. For the scattering and the bound state parts, according to the selection criterion given in Sec.~\ref{ComputEoS}, all possible color channels  and all helicity channels with $\left\langle \bm L^2\right\rangle \le 5$ (except $\left|S_-;1^{+} \{4\}\right\rangle$ and $\left|D_-;2^{-} \{4\}\right\rangle$ $gg$ states) are included in the computation (see Appendices~\ref{colorchan} and \ref{helicity}).

\begin{figure}[h!]
\begin{center}
\includegraphics*[width=0.7\textwidth]{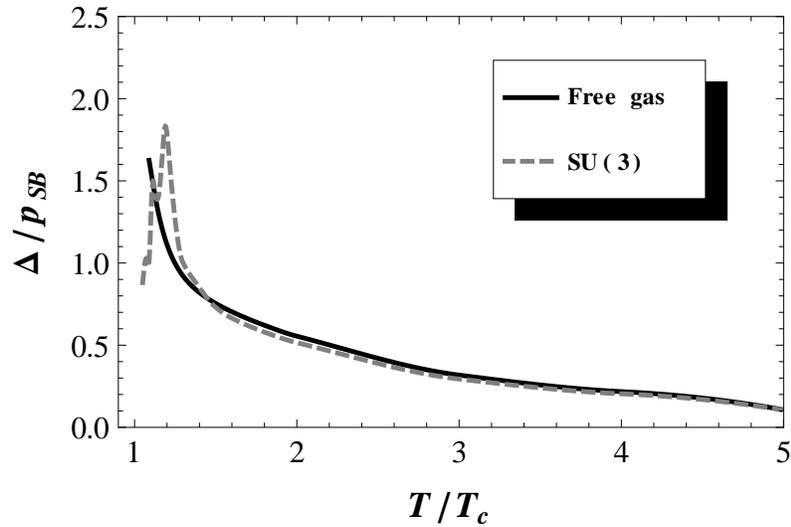}
\end{center}
\caption{Normalized trace anomaly $\Delta/p_{SB}$ versus temperature in units of $T_c$, compared to the normalized free gas part of $\Omega$, $\Omega_0$.}
\label{anom_Susy}
\end{figure}

The global structure has the same features as previously observed in the ordinary YM case \cite{LSCB}. The bound-state contribution is weak at every computed temperature but it is logically maximal around 1.2 $T_c$, where the trace anomaly appears to be maximal (see Fig.~\ref{anom_Susy}). The scattering part is also maximal around 1.2 $T_c$. At higher temperatures the repulsive $gg$ channel $(2,0,\dots,0,2)$ is dominant in the scattering part and tends to decrease the free gas part. Nevertheless, the effect of two-body interactions on the pressure is a minor contribution to the total pressure. This validates the use of the virial expansion~(\ref{pot0}) for which deviations from the ideal gas of massive particles must be weak. The slight differences between the ordinary YM and the ${\cal N} = 1$ SUSY YM are: The bound state sector is richer since $\tilde{g}\tilde{g}$, and $g\tilde{g}$ bound states exist up to 1.3 $T_c$ in their most attractive channel, and the scattering part is naturally build with much more channels. 

\begin{figure}[h!]
\begin{center}
\includegraphics*[width=0.49\textwidth]{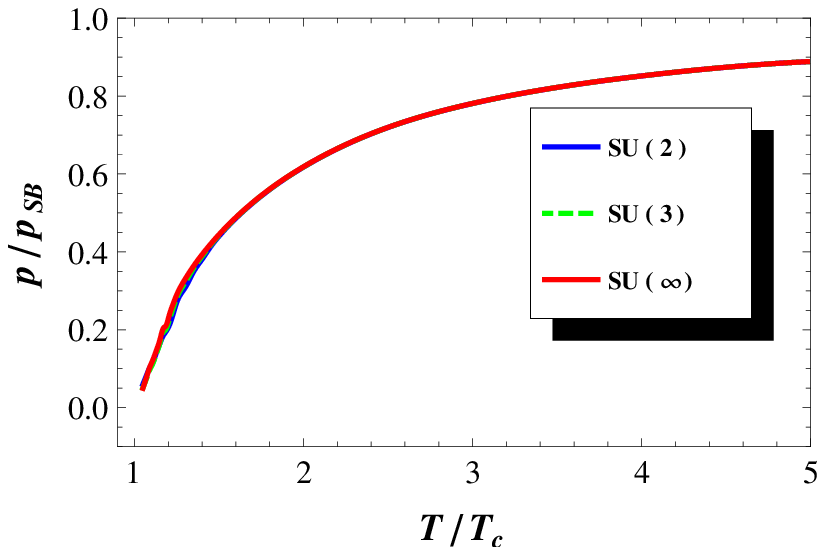}
\includegraphics*[width=0.49\textwidth]{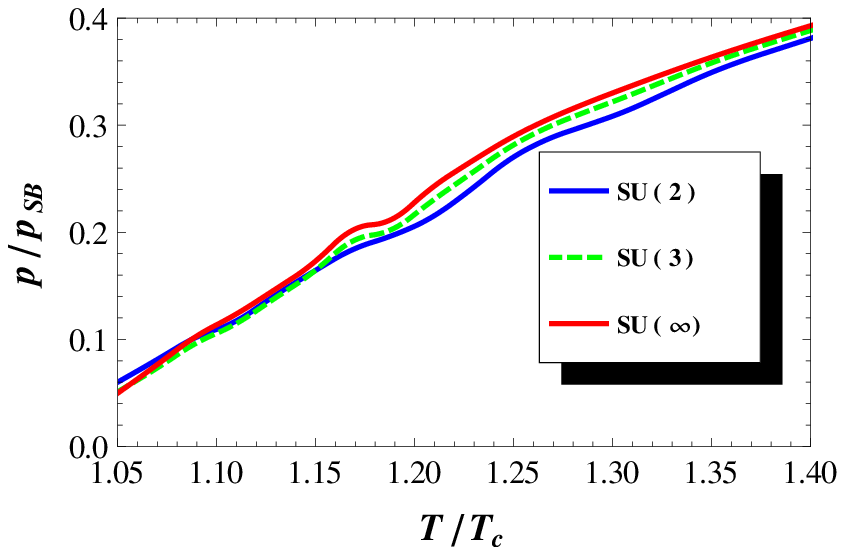}
\end{center}
\caption{(color online) (Left) Normalized pressure $p/p_{SB}$ versus temperature in units of $T_c$, computed for different SU($N$) gauge groups. (Right) Zoom near $T_c$. }
\label{zoom_Susy}
\end{figure}

\begin{figure}[h!]
\begin{center}
\includegraphics*[width=0.7\textwidth]{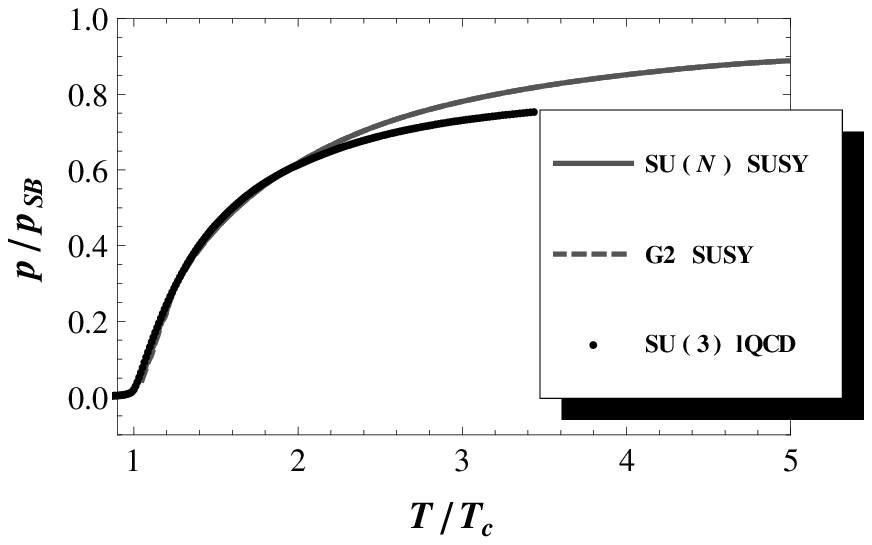}
\caption{Normalized pressures $p/p_{SB}$ versus $T/T_c$, with gauge groups SU(2), SU(3) and SU($\infty$) commonly denoted as SU($N$) (solid gray line) and G$_2$ (dashed gray line). The pressure for non-SUSY YM with gauge group SU(3) is shown for comparison (black dots); data are taken from the lattice study \cite{panero}.}
\label{pressure}
\end{center}
\end{figure}

In Fig.~\ref{zoom_Susy}, the SU($N$) gauge structure of the normalized total pressure is shown. Again in this case, an asymptotically SU($N$) gauge-group independent observable seems to emerge: The maximum deviation between the curves is just above $T_c$. This is in agreement with the scaling relations introduced in \cite{LSCB}. Indeed, if we expand the $T$-matrix in terms of $V$, the color dependence of the total pressure (without considering bound states) is in dim $adj$ up to $O(V^3)$. This factor is thus cancelled by the same one in the pressure of normalization, and so the gauge-group independence is obtained at high temperature since such expansion of the $T$-matrix becomes more and more valid with an increase of $T$.   Moreover, it is worth adding that this extraction of the color factors is possible only because the gluon and gluino mass is gauge-group independent in our framework. The normalized pressure with gauge group $G_2$ is displayed in Fig. \ref{pressure}; the curve is nearly indistinguishable from the SU($N$) case.

\begin{figure}[h!]
\begin{center}
\includegraphics*[width=0.7\textwidth]{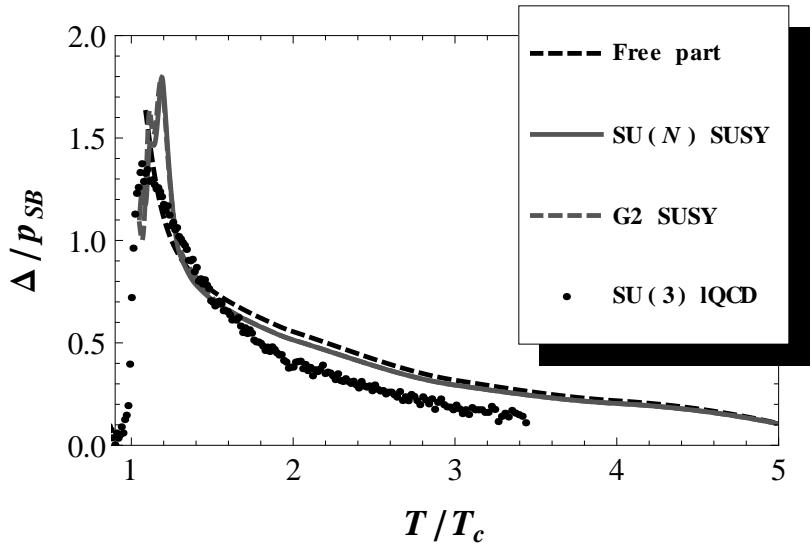}
\end{center}
\caption{Same as Fig.~\ref{pressure} for the normalized trace anomaly $\Delta/p_{SB}$. The free part of the trace anomaly is shown for comparison (dashed black line) and is gauge-group independent.  Lattice data of \cite{panero} corresponding to the ordinary SU(3) YM case are also indicated for comparison (dots).}
\label{traceanom}
\end{figure}

The behavior of the total normalized trace anomaly is displayed in Fig.~\ref{anom_Susy}. Again, we find appropriate to compute it without bound state (the treatment of bound states requires a more refined study \cite{LSCB}) and to compare it to the normalized free gas part $\Omega_0$. We also observe in this case that the interactions provide the peak structure and that the trace anomaly tends to zero at high $T$; the free part alone would not lead to this peak structure. The normalized trace anomaly for SU($N$) and $G_2$ groups is presented in Fig.~\ref{traceanom}. The asymptotic gauge-group universality is, without surprise, observed and follows the same justification as the normalized pressure. The behavior around $T_c$ exhibits some slight differences according to the gauge group. These deviations are discussed in the next section.

Note that the ordinary and SUSY YM EoS are not predicted to be drastically different, as shown in Fig. \ref{traceanom} through a comparison to the non-SUSY lattice EoS of \cite{panero}.  



\subsection{Influence of the gauge group}
Although we focused on SU($N$) and G$_2$ in our numerical calculations, some universal features can be expected at the level of the EoS. First, one has $adj\otimes adj=\bullet^S\oplus adj^A\oplus \textrm{higher\ dim}$, for all gauge groups. The singlet generates bound states and attractive interactions with $\kappa_\bullet=-1$, but its contribution to the EoS is subleading. The adjoint representation also leads to bound states and attractive interactions (with $\kappa_{adj}=-1/2$), but in this case the contributions to the EoS is dominant since it scales in $\mathrm{dim}\, adj$. Information about the higher dimensional representations can be obtained from Eqs. (\ref{eqS}) and (\ref{eqA}):
\begin{equation}\label{univ}
\sum_{{\cal C}_S} {\rm dim}{\cal C}_S\, \kappa_{{\cal C}_S}=\frac{\mathrm{dim}\, adj}{2} , \quad 
\sum_{{\cal C}_A} {\rm dim}{\cal C}_A\, \kappa_{{\cal C}_A}=-\frac{\mathrm{dim}\, adj}{2}.
\end{equation}
Because the singlet is symmetric and with $\kappa_\bullet<0$, there exists at least one color channel generating repulsive interactions for any gauge group. Moreover, $\sum_{{\cal C}_A\neq adj} {\rm dim}{\cal C}_A\, \kappa_{{\cal C}_A}=0$. It can be checked by explicit computation that only two antisymmetric channels are actually present in $adj \otimes adj$: the adjoint and another channel leading to vanishing two-body interactions in agreement with this last equality. So the number of color channels and the value of the associated color factors may differ from one gauge group to another, but qualitatively influence the EoS in the same way. 

The differences between different gauge groups appear more clearly by noticing that the on-shell $\mathcal{T}$-matrix can be written as ${\cal T}= \sum_k a_k\,\kappa_{{\cal C}}^k $ where all $a_k$ do not depend on the color but rather on the other quantum numbers involved. The color dependence of the thermodynamic observables is then driven by the quantities $\sum_{{\cal C}_{A/S}} \text{dim} \, {\cal C}_{A/S}\,\kappa_{{\cal C}_{A/S}}^k$. Equations (\ref{univ}) show that the EoS normalized to the SB pressure is gauge-group-independent at order $k=1$, which is the Born approximation. By explicit computation for all gauge groups, it can be checked that the differences only appear at order $k=4$. For example, $\sum_{{\cal C}_{S}} \text{dim} \, {\cal C}_{S}\,\kappa_{{\cal C}_{S}}^4$ is no longer proportionnal to $N^2-1$ for SU($N$), so the contribution to the normalized EoS is different for each $N$ at this order.

The normalized EoS above deconfinement is thus predicted to be very weakly gauge-group dependent in our approach, for SUSY and non-SUSY YM. Such a result is in agreement with the recent lattice study \cite{Bruno} showing that the SU($N$) and $G_2$ normalized EoS are indeed compatible above $T_c$.

\section{High temperatures}\label{highT}

The interaction potential progressively vanishes at high temperatures because of the increasing screening. Consequently the high temperature limit of our model should be accurately described within the Born approximation ${\cal T}=V+\,{\rm O}(V^2)$. It has been shown in \cite{LSCB} that the interactions between two different species vanish within this approximation because of an identity relating the color factors: $\sum_{{\cal C}}{\rm dim}\,{\cal C} \ \kappa_{{\cal C},ij}=0$. This sum actually appears when summing the different color contribution to the grand potential of a given channel involving two different species. When two identical species are involved, this argument does not hold because the summation is restricted to channels with a given symmetry.

\begin{figure}[h!]
\begin{center}
\includegraphics*[width=0.7\textwidth]{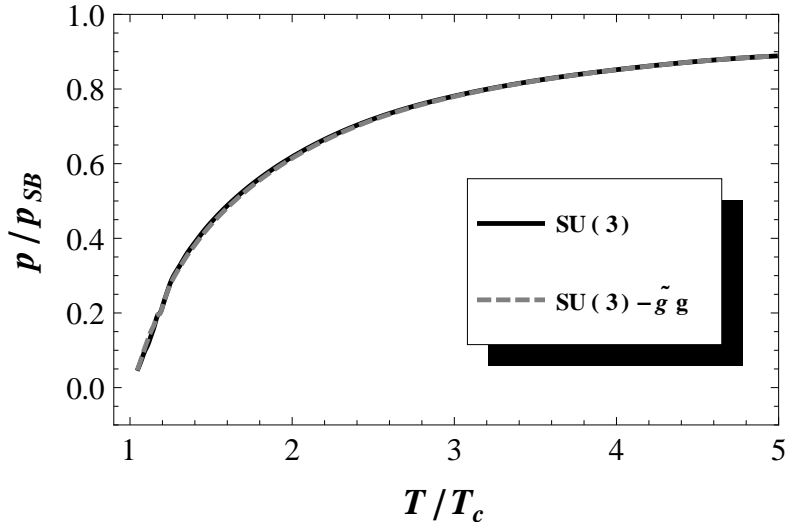}
\end{center}
\caption{Normalized pressure $p/p_{SB}$ versus temperature in units of $T_c$ for the gauge group SU(3) in the ${\cal N}=1$ SUSY YM case: All contributions (solid line) and  all contributions except $\tilde g g$ channels (dashed line).}
\label{susywithoutqg}
\end{figure}

In conclusion, gluons and gluinos do not interact with each other at high temperature in average. However, the interactions between gluons only and gluinos only are still present. We have checked that taking into account or not the contributions of $\tilde g g$ pairs make no noticeable difference for the pressure. Curves with and without these contributions are indistinguishable (see Fig.~\ref{susywithoutqg}), except just above $T_c$ where the pressure is very slightly increased without the contributions of $\tilde g g$ channels.

It is worth mentioning that the thermal masses we use here are based on fits of lattice data which are valid between typically (1-4) $T_c$. At very high temperatures ($T\gg T_c$), the thermal masses should be equal to what is obtained within the Hard-Thermal-Loop (HTL) formalism and the EoS of the medium would thus be given by a free gas of particles with HTL thermal masses, assuming $V=0$. Such a model can be found in \cite{Minimal}, where useful references about HTL results are given. Note finally that reaching pressure compatible with $p_{SB}$ demands huge temperatures: around $10^7$ $T_c$ as shown in \cite{highT}.

\section{Orientifold equivalence}\label{qcdasp}

Let us denote by QCD$_{AS}$ a SU($N$) YM theory with $N_f$  Dirac fermions in the two-index antisymmetric representation $(0,1,0,\dots,0)$, and QCD$_{adj}$ an SU($N$) YM theory with $N_f$ Majorana flavors in the adjoint representation. The so-called orientifold equivalence states that QCD$_{AS}$ and QCD$_{adj}$ are equivalent at large $N$ in the bosonic sector \cite{Armo03}. This equivalence is particularly appealing when $N_f = 1$ since in this case QCD$_{adj}$ is actually $\mathcal{N} = 1$ SUSY YM. Moreover, QCD$_{AS}$ reduces to standard one-flavor QCD ($I=0$) for $N=3$. 

Within our framework it is possible to show that orientifold equivalence holds, and to compute how far one-flavor QCD deviates for the large-$N$ limit. This is the purpose of the present section where we notice that the meaning of the symbol $\cong$ will be ``equal at the limit $N\rightarrow\infty$". Moreover, $q_A$ ($\bar q_A$) will denote a(n) (anti)quark in the two-index antisymmetric representation. Possible color channels pairs containing $q_A$ or $\bar q_A$ are given in Appendix~\ref{colorchan}, and helicity channels in Appendix~\ref{helicity}. The various pairs to be considered are $g g$, $g Q$ and $QQ$, where $Q$ is a generic term for $q_A$ and $\bar q_A$. The approximations assumed for the interactions taken into account are of the same kind that the ones considered for the plasma $g \tilde g$, $g \tilde g$ and $\tilde g \tilde g$: Strict respect of the Casimir scaling and no annihilation contributions. 

First, we have to check that the masses of the particles coincide. The gluon thermal mass is common to QCD$_{AS}$ and QCD$_{adj}$ since the gluonic sector is identical in both theories. An assumption of our model is that the function $\Delta(T)$ is gauge-group independent. Hence, $\delta_{\tilde g}(T)=\delta_g(T)=\Delta(T)$. Using the color factors listed in Appendix \ref{colorchan}, one gets
\begin{equation}
\delta_{q_A}(T) =\delta_{\bar q_A}(T)= \sqrt{\frac{N^2 - N -2}{N^2}} \Delta(T)
\end{equation}
which is not equal to $\Delta (T)$ in general. Nevertheless, $\delta_{q_A}(T) =\delta_{\bar q_A}(T)\cong\delta_{\tilde g}(T)$ as expected. To obtain the equality between the thermal masses $m_{\tilde g}$, $m_{q_A}$ and $m_{\bar q_A}$ at large $N$, it is necessary to take the same value of the parameter $m_0$ for the three particles (a study with physical quark masses will be presented elsewhere \cite{prepa}). As $m_0$ is assumed to be gauge-group-independent, this parameter is constant with $N$. The free gluinos thus bring a contribution $\Omega_0(\tilde g)=2 (N^2-1) \omega_-(m_{\tilde g})$ to the grand potential, while the free quarks and antiquarks bring a corresponding contribution $\Omega_0(q_A,\bar q_A)=2\frac{N(N-1)}{2}\omega_-(m_{q_A})+2\frac{N(N-1)}{2}\omega_-(m_{\bar q_A})$. It is then straightforwardly checked that 
\begin{equation}
\Omega_0(\tilde g)\cong\Omega_0(q_A,\bar q_A).
\end{equation}
After the equivalence of the free part, we have to show the equivalence of the two-body contributions. The $gg$ channels are trivially equal in  QCD$_{AS}$ and QCD$_{adj}$, so only the $\tilde g\tilde g$ and $\tilde g g$ channels have to be investigated. 

The $\tilde g\tilde g$ channels are bosonic; their contribution should thus be equivalent to that of the $q_A q_A$, $\bar q_A q_A$ and $\bar q_A\bar q_A$ ones. The color singlet appears in $\tilde g\tilde g$ and $q_A\bar q_A$; the corresponding contributions to the pressure are different because the Pauli principle has to be applied in the first one, not in the second one. This discrepancy is irrelevant at large $N$ because the singlet's contribution is of order 1, not $N^2$. The symmetric and antisymmetric adjoint channels in $\tilde g\tilde g$ bring a pressure contribution which is equal to the adjoint channel appearing in $q_A\bar q_A$: All the possible helicity states are allowed in both cases. The $(2,0,\dots,1,0)$ and $(0,1,\dots,0,2)$ channels in $\tilde g\tilde g$ have no equivalent in the quark case, but they bring no contribution to the EoS since $\kappa_{\cal C}=0$. The remaining color channels in $\tilde g\tilde g$ are the symmetric $(2,0,\dots,0,2)$ and $(0,1,\dots, 1,0)$ ones, that should match with the symmetric $(0,0,0,1,0\dots,0)$ and $(0, \dots,0,1,0,0,0)$ ones in $q_A q_A$ and $\bar q_A\bar q_A$. This is actually the case. Indeed, Pauli principle asks the $\tilde g\tilde g$ states to have $L+S$ even, just as for the $q_A q_A$ and $\bar q_A\bar q_A$ states. Moreover, the Born approximation is valid at large $N$ for all those channels since $\kappa_{\cal C}=O(1/N)$. The contributions of the $\tilde g\tilde g$ channels to the grand potential is thus proportional to $\sum_{{\cal C}} {\rm dim}\ {{\cal C}}\ \kappa_{{\cal C}}= (N^2-1)/2$, as well as the $q_Aq_A$, $\bar q_A\bar q_A$, $q_A\bar q_A$ ones for which $\sum_{{\cal C}} {\rm dim}\ {{\cal C}}\ \kappa_{{\cal C}}+\sum_{\bar {\cal C}} {\rm dim}\ {\bar {\cal C}}\ \kappa_{\bar {\cal C}}= (N^2-1)(N-2)/(2N)$. Both factors are equal at large $N$, leading to equivalent contributions to the EoS. 

Finally, the $\tilde g g$ contribution should be equivalent to the $q_A g$ and $\bar q_A g$ ones. The same kind of arguments apply, so we will not perform the full analysis for the sake of clarity. Let us just mention that the two adjoint channels in $\tilde g g$ bring equivalent contributions to the grand potential than the $(2,0,\dots,0)$ and $(0,1,0,\dots,0)$ channels in $q_A g$ an $\bar q_A g$. Similarly, the $(2,0,\dots,0,2)$ and $(0,1,\dots, 1,0)$ channels in $\tilde g g$ match with the $(1,1,0,\dots,0,1)$ and $(0,0,1,0,\dots,0,1)$ one in $q_A g$ an $\bar q_A g$. 

\begin{figure}[h!]
\begin{center}
\includegraphics*[width=0.45\textwidth]{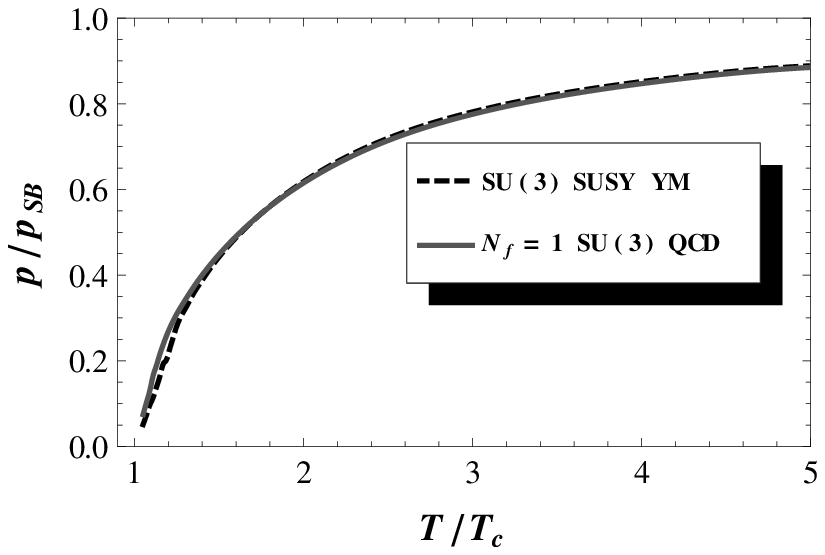}
\includegraphics*[width=0.45\textwidth]{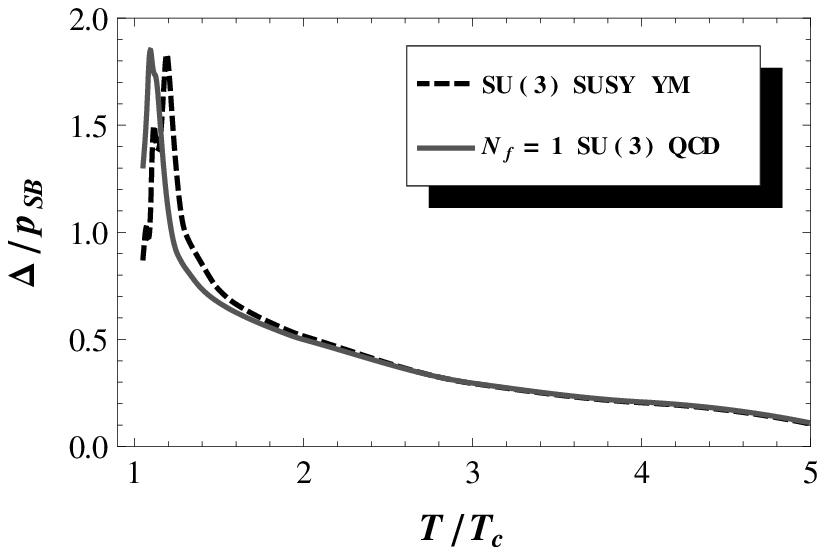}
\caption{(Left) Normalized pressure $p/p_{SB}$ versus temperature in units of $T_c$, computed for SU(3) QCD with $N_f = 1$ and ${\cal N}=1$ SU(3) SUSY YM. (Right) Normalized trace anomaly $\Delta/p_{SB}$ versus temperature in units of $T_c$, computed for SU(3) QCD with $N_f = 1$  and ${\cal N}=1$ SU(3) SUSY YM. 
}
\label{Orient1}
\end{center}
\end{figure}

The above discussion shows that $\textrm{QCD}_{AS}\cong \textrm{QCD}_{adj}$: The orientifold equivalence is checked within our framework. This can be seen as a strong validation of the various assumptions made in the building of the model. In particular, the Casimir scaling hypothesis for the interaction is perfectly compatible with this equivalence. Now we can compare the accuracy of the orientifold equivalence at finite $N$, namely $N=3$. The pressure and trace anomaly of SU(3) ${\cal N}=1$ SUSY YM is compared to the EoS of SU(3) one-flavor QCD in Fig. \ref{Orient1}. 
The free part is given by, for SU($N$) gauge groups,
\begin{equation}\label{Om0qa}
\Omega_0 = 2 (N^2-1) \omega_+(m_g) + 2 N (N-1) \omega_-(m_{q_A}).
\end{equation}
For the scattering and the bound state parts, according to the selection criterion given in Appendix~\ref{cross}, all possible color channels and all helicity channels with $\left\langle \bm L^2\right\rangle \le 5$ (except $\left|S_-;1^{+} \{4\}\right\rangle$ and $\left|D_-;2^{-} \{4\}\right\rangle$ $gg$ states) are included in the computation. These plots show how far one-flavor QCD is from the ${\cal N}=1$ SUSY YM theory at the level of the EoS. As far as the pressure is concerned, both theories are very similar. The trace anomaly however reveals some differences around 1.2 $T_c$. Note that each case is normalized to its own Stefan-Boltzmann pressure, that of QCD$_{AS}$ reading $p_{SB}=\pi^2 T^4 (N-1)(\frac{15}{8}N+1)/45$ for SU($N$) gauge groups.

\section{Conclusions}\label{conclu}

We have studied the properties of ${\cal N}=1$ SUSY YM within quasiparticle approaches, with an emphasis on a $T$-matrix formulation that had already proven to be successful in the modeling of ordinary YM theory in the deconfined phase. Here is a summary of the results we have found: 

\begin{itemize}
\item  Among the various possible bound states, the $a-\eta'$, the scalar glueball and the lightest gluino-gluon bound state can survive up to 1.4 $T_c$, implying that the gluon-gluino plasma is rich in bound states in the temperature range where the trace anomaly is maximal. Its non-ideality is thus clearly related to the strength of the interactions, which are strong enough to create bound states. 
\item The bound-state properties are independent of the gauge group in the singlet channel, and the normalized EoS is only weakly-dependent on the gauge group. 
\item At large temperatures, the trace anomaly decreases as expected from the $\beta$-function, and the two-body gluon-gluino interactions are suppressed while it is not the case for the gluon-gluon and gluino-gluino interactions. 
\item The orientifold duality holds at the level of the EoS within our formalism. Some differences exist between one-flavor QCD and SU(3) ${\cal N}=1$ SUSY YM at the level of the trace anomaly. Such a disagreement is not unexpected since the duality is exact at large-$N$ only.  
\end{itemize}

We have focused here on SU($N$) and $G_2$ gauge groups for the computation of the EoS. However, our formalism can be implemented for any gauge group. The next step would be the computation of transport coefficients with and without SUSY. The problem is worth to be studied and is left for future works, as well as the extension of our formalism to a larger number of supersymmetries. 

\textit{Acknowledgments --}
G. L. thanks the F.R.S-FNRS for financial support. The authors thank D. Cabrera for useful remarks and suggestions.

\appendix

\section{Two-body color channels for SU($N$) and $G_2$} \label{colorchan}

Characteristics of color channels for particles in the adjoint representation of SU($N$) can be found in \cite{LSCB}, but they are recalled here in Table~\ref{tab:c0} for completeness: condition of existence as a function of $N$, possible symmetry (Symmetrical or Antisymmetrical), dimension and color factor ($\kappa_{{\cal C}}$ defined in (\ref{color_scaling})). Characteristics of color channels, implying a particle in the antisymmetrical $(0,1,0, \ldots,0)$ representation of SU($N$), are given in Table~\ref{tab:c1}-\ref{tab:c3}. Data for the $G_2$ group are given in Table~\ref{tabg2}. One can check that $\sum_{{\cal C}}{\rm dim}\,{\cal C}= {\rm dim}\, R_1 \times {\rm dim}\, R_2$ and $\sum_{{\cal C}}{\rm dim}\,{\cal C} \ \kappa_{{\cal C},12}=0$. Note that a general method for computing the quadratic Casimir of SU($N$) can be found in \cite{cas}. 

\begin{table}[ht]\caption{Condition of existence, symmetry, dimension and color factor ($\kappa_{{\cal C}}$) of the color channels (${\cal C}$) appearing in the tensor products of $g\,g$ channels, where $g$ is the adjoint representation of SU($N$).}
\begin{center}
\begin{tabular}{cccccc}
\hline\hline       
${\cal C}$ & $\bullet$ & $(1,0,\dots,0,1)$ & $(2,0,\dots,0,2)$ & $(2,0,\dots,1,0)$    & $(0,1,0,\dots,0,1,0)$ \\
&             &                 &                 & $(0,1,\dots,2)$  &                     \\
\hline
$N\ge$ & 2 & 3(S),2(A) & 2 & 3 & 4 \\
Symmetry & S & S,A & S & A & S  \\
Dimension & \scriptsize{1}  & \scriptsize{$N^ 2-1$} &$\frac{N^2(N+3)(N-1)}{4}$ & $\frac{(N^ 2-4)(N^ 2-1)}{4}$ & $\frac{N^2(N-3)(N+1)}{4}$ \\     
$\kappa_{{\cal C}}$ & \scriptsize{$-1$} & $-\frac{1}{2}$ & $\frac{1}{N}$ & \scriptsize{0} & $-\frac{1}{N}$ \\
\hline\hline
\end{tabular}
\end{center}
\label{tab:c0}
\end{table}

\begin{table}[htb]\caption{Same as Table~\ref{tab:c0}, but for the tensor products of $q_A\, q_A$, where $q_A$ is the antisymmetrical $(0,1,0, \ldots,0)$ representation of SU($N$) with dimension $N(N-1)/2$. For each $q_A\, q_A$ channel ${\cal C}$, a $\bar q_A\, \bar q_A$ channel ${\cal \bar C}$ exists with the same characteristics.} 
\begin{center}
\begin{tabular}{cccc}
\hline\hline
${\cal C}$ & $(0,2,0,\dots,0)$ & $(1,0,1,0\dots,0)$ & $(0,0,0,1,0\dots,0)$ \\
${\cal \bar C}$ & $(0,\dots,0,2,0)$ & $(0,\dots,0,1,0,1)$ & $(0,\dots,0,1,0,0,0)$ \\
\hline
$N\ge$ & 2 & 3 & 4 \\
Symmetry & S & A & S \\
Dimension & $\frac{N^2(N^2-1)}{12}$ & $\frac{N(N^2-1)(N-2)}{8}$ & $\frac{N(N-1)(N-2)(N-3)}{24}$ \\     
$\kappa_{{\cal C}}$ & $\frac{N-2}{N^2}$ & $-\frac{2}{N^2}$ & $-\frac{2(N+1)}{N^2}$ \\
\hline\hline
\end{tabular}
\end{center}
\label{tab:c1}
\end{table}

\begin{table}[htb]\caption{Same as Table~\ref{tab:c0}, but for $q_A\, \bar q_A$ channels.} 
\begin{center}
\begin{tabular}{cccc}
\hline\hline
${\cal C}$ & $\bullet$ & $(1,0,\dots,0,1)$ & $(0,1,0, \ldots,0,1,0) $\\
\hline
$N\ge$ & 2 & 3 & 4 \\
Dimension & \scriptsize{1} & \scriptsize{$N^2-1$} & $\frac{(N-3)N^2(N+1)}{4}$  \\     
$\kappa_{{\cal C}}$ & $-\frac{(N-2)(N+1)}{N^2}$ & $\frac{-N^2+2N+4}{2N^2}$ & $\frac{2}{N^2}$  \\
\hline\hline
\end{tabular}
\end{center}
\label{tab:c2}
\end{table}

\begin{table}[htb]\caption{Same as Table~\ref{tab:c0}, but for $q_A\,g$ channels. For each $q_A\, g$ channel ${\cal C}$, a $\bar q_A\, g$ channel ${\cal \bar C}$ exists with the same characteristics.} 
\begin{center}
\begin{tabular}{ccccc}
\hline\hline
${\cal C}$ & $(2,0,\dots,0)$ & $(0,1,0\dots,0)$ & $(1,1,0,\dots,0,1)$ &  $(0,0,1,0,\ldots,0,1)$ \\
${\cal \bar C}$ & $(0,\dots,0,2)$ & $(0,\dots,0,1,0)$ & $(1,0,\dots,0,1,1)$ &  $(1,0,\ldots,0,1,0,0)$ \\
\hline
$N\ge$ & 2 & 3 & 3 & 4 \\
Dimension & $\frac{N(N+1)}{2}$ & $\frac{N(N-1)}{2}$ & $\frac{N^2(N^2-4)}{3}$ & 
$\frac{N(N^2-1)(N-3)}{6}$ \\     
$\kappa_{{\cal C}}$ & $-\frac{N-2}{2N}$ & $-\frac{1}{2}$ & $\frac{1}{2N}$ & $-\frac{1}{N}$\\
\hline\hline
\end{tabular}
\end{center}
\label{tab:c3}
\end{table}

\begin{table}[ht]\caption{Same as Table~\ref{tab:c0}, but for the group $G_2$.}
\begin{center}
\begin{tabular}{cccccc}
\hline\hline       
${\cal C}$ & $\bullet$ & $(0,1)$ & $(0,2)$ & $(2,0)$    & $(3,0)$ \\
\hline
Symmetry & S & A & S & S & A  \\
Dimension & \scriptsize{1}  & \scriptsize{14} &$77$ & $27$ & $77$ \\     
$\kappa_{{\cal C}}$ & \scriptsize{$-1$} & $-\frac{1}{2}$ & $\frac{1}{4}$ &  $-\frac{5}{12}$& \scriptsize{0} \\
\hline\hline
\end{tabular}
\end{center}
\label{tabg2}
\end{table}

\section{Helicity formalism for spin-1/2 and transverse spin-1 particles} \label{helicity}

The proper way to manage two-body states containing gluons, gluinos or quarks is to use the Jacob and Wick's helicity formalism~\cite{jaco}, since a gluon is a transverse spin-1 particle and a quark or a gluino is a spin-1/2 particle. A two-body state with total spin $J$, with helicities $\lambda_1$ and $\lambda_2$, and with a given parity $P$ can be written
\begin{equation}\label{hstate}
\left|J^P,M;\lambda_1,\lambda_2,\epsilon\right\rangle=\frac{1}{\sqrt 2} \left[ \left|J,M;\lambda_1,\lambda_2\right\rangle+\epsilon\left|J,M;-\lambda_1,-\lambda_2\right\rangle\right],
\end{equation}
with $\epsilon=\pm1$ and $\left|J,M;\lambda_1,\lambda_2\right\rangle$ a two-particle helicity state in the rest frame of the system. The parity is given by $P=\epsilon\, \eta_1\eta_2(-1)^{J-s_1-s_2}$ with $\eta_i$ and $s_i$ the intrinsic parity and spin of particle $i$. Moreover, $J\geq|\lambda_1-\lambda_2|$. The helicity states can be expressed as particular linear combinations of usual normalized basis states $\left|^{2S+1}L_J\right\rangle$, which is very convenient to perform the computations \cite{jaco}
\begin{eqnarray}\label{decomp}
    \left|J,M;\lambda_1,\lambda_2\right\rangle&=&\sum_{L,S}\left[\frac{2L+1}{2J+1}\right]^{1/2} 
\langle L,S;0,\lambda_1-\lambda_2|J,\lambda_1-\lambda_2\rangle \nonumber \\
&&\times \langle s_1,s_2;\lambda_1,-\lambda_2|S,\lambda_1-\lambda_2\rangle\, \left|^{2S+1}L_J\right\rangle.
\end{eqnarray}

For the present work, it is sufficient to recall that four families of helicity states can be found, separated in helicity singlets 
\begin{equation}\label{Spm}
\left|S_\pm; J^P\right\rangle = 
\left|J^P,M;\lambda_1,\lambda_2,\pm\right\rangle \quad \textrm{with} \quad
\lambda_1 \lambda_2 > 0 
\end{equation}
and doublets  
\begin{equation}\label{Dpm}
\left|D_\pm; J^P\right\rangle =
\left|J^P,M;\lambda_1,\lambda_2,\pm\right\rangle \quad \textrm{with} \quad
\lambda_1 \lambda_2 < 0.
\end{equation}
This notation follows the pioneering work~\cite{barnes} and is used in \cite{heli}. In the special case of identical particles, the helicity states must also be eigenstates of the operator $\hat {\cal S}=\left[1+(-1)^{2s}P_{12}\right]$, which is the projector on the symmetric ($s$ integer) or antisymmetric ($s$ half-integer) part of the helicity state. It can be seen that the states \cite{heli}
\begin{equation}\label{hdef}
\left|J^P,M;\lambda_1,\lambda_2,\epsilon,\rho\right\rangle=\frac{1}{2}\left\{\left|J^P,M;\lambda_1,\lambda_2,\epsilon\right\rangle+\rho\left|J^P,M;\lambda_2,\lambda_1,\epsilon\right\rangle\right\}, 
\end{equation}
with $\rho=\pm 1$, are eigenstates of $\hat{\cal S}$ with the eigenvalues $1+\rho(-1)^J$.

\subsection{Two transverse spin-1 particles}

The general forms of the two-gluon states have been presented in \cite{heli}. Some properties are given in Table~\ref{tab:heli1}, as well as the average value of some operators, computed with these states. For completeness, the ones considered in this paper are recalled here. 

\begin{table}[htb]
\caption{Symmetrized (S) and antisymmetrized (A) two-gluon helicity states, following the notation of~\cite{barnes,heli}, with the corresponding quantum numbers and some averaged operators.} 
\begin{center}
\begin{tabular}{cccccc}
\hline\hline
State & S & A & $\left\langle \bm L^2\right\rangle$ & $\left\langle \bm S^2\right\rangle$ & $\left\langle \bm L\cdot \bm S\right\rangle$\\
\hline
$\left|S_+; J^P\right\rangle$ & (even-$J\geq0$)$^+$ & (odd-$J\geq1$)$^-$ & $J(J+1)+2$ & 2 & $-2$ \\
$\left|S_-; J^P\right\rangle$ & (even-$J\geq0$)$^-$ & (odd-$J\geq1$)$^+$ & $J(J+1)+2$ & 2 & $-2$  \\
$\left|D_+; J^P\right\rangle$ & (even-$J\geq2$)$^+$ & (odd-$J\geq3$)$^-$ & $J(J+1)-2$ & 6 & $-2$  \\
$\left|D_-; J^P\right\rangle$ & (odd-$J\geq3$)$^+$  & (even-$J\geq2$)$^-$ & $J(J+1)-2$ & 6 & $-2$  \\
\hline\hline
\end{tabular}
\end{center}
\label{tab:heli1}
\end{table}

The number between braces being the value of $\left\langle \bm L^2\right\rangle$, the first symmetric states are:
\begin{eqnarray}\label{scaps}
    \left|S_+;0^{+} \{2\}\right\rangle&=&\left[\frac{2}{3}\right]^{1/2}\left|^1 S_0\right\rangle+\left[\frac{1}{3}\right]^{1/2}\left|^5 D_0\right\rangle,\\
    \left|S_-;0^{-} \{2\}\right\rangle&=&-\left|^3P_0\right\rangle, \\
    \left|D_+;2^{+} \{4\}\right\rangle&=&\left[\frac{2}{5}\right]^{1/2}\left|^5S_2\right\rangle
    +\left[\frac{4}{7}\right]^{1/2}\left|^5 D_2\right\rangle+\left[\frac{1}{35}\right]^{1/2}\left|^5 G_2\right\rangle.
\end{eqnarray}    
The first antisymmetric states are:
\begin{eqnarray}
    \left|S_+;1^{-} \{4\}\right\rangle&=&\left[\frac{2}{3}\right]^{1/2}\left|^1 P_1\right\rangle-\left[\frac{2}{15}\right]^{1/2}\left|^5 P_1\right\rangle+\left[\frac{1}{5}\right]^{1/2}\left|^5 F_1\right\rangle , \\ 
    \left|S_-;1^{+} \{4\}\right\rangle&=&\left[\frac{1}{3}\right]^{1/2}\left|^3 S_1\right\rangle-\left[\frac{2}{3}\right]^{1/2}\left|^3 D_1\right\rangle ,\\
     \left|D_-;2^{-} \{4\}\right\rangle&=&-\left[\frac{4}{5}\right]^{1/2}\left|^5 P_2\right\rangle-\left[\frac{1}{5}\right]^{1/2}\left|^5 F_2\right\rangle .
\end{eqnarray}
All other states are characterized by $\left\langle \bm L^2\right\rangle \ge 8$; their general form can be found in \cite{heli}. 

\subsection{States containing one transverse spin-1 particle}

States containing a gluon and a gluino or a quark can also be expressed as particular linear combinations of usual basis states $\left|^{2S+1}L_J\right\rangle$, following the procedure given in \cite{heli}. Some properties are given in Table~\ref{tab:heli2}, as well as the average value of some operators, computed with these states.

\begin{table}[htb]
\caption{Helicity states containing a gluon and a gluino or a quark, following the notation of~\cite{barnes,heli}, with the corresponding quantum numbers and some averaged operators.} 
\begin{center}
\begin{tabular}{ccccc}
\hline\hline
State & $J$ min & $\left\langle \bm L^2\right\rangle$ & $\left\langle \bm S^2\right\rangle$ & $\left\langle \bm L\cdot \bm S\right\rangle$\\
\hline
$\left|S_\pm; J^P\right\rangle$ & $\frac{1}{2}$ & $J(J+1)+\frac{5}{4}$ & $\frac{7}{4}$ & $-\frac{3}{2}$ \\
$\left|D_\pm; J^P\right\rangle$ & $\frac{3}{2}$ & $J(J+1)-\frac{3}{4}$ & $\frac{15}{4}$ & $-\frac{3}{2}$  \\
\hline\hline
\end{tabular}
\end{center}
\label{tab:heli2}
\end{table}

The general forms of the $q\,g$ states read as
\begin{eqnarray}
\left|S_+;J^{(-1)^{j-1/2}}\right\rangle & = & \left[\frac{2J-1}{8J}\right]^{1/2}\left|^4 J-3/2_{J} \right\rangle 
+\left[\frac{2}{3}\right]^{1/2}\left|^2 J+1/2_{J} \right\rangle \nonumber \\
&&-\left[\frac{2J+3}{24J}\right]^{1/2}\left|^4 J+1/2_{J} \right\rangle, \\
\left|S_-;J^{(-1)^{j+1/2}}\right\rangle & = & \left[\frac{2}{3}\right]^{1/2}\left|^2 J-1/2_{J} \right\rangle 
+\left[\frac{2J-1}{24(J+1)}\right]^{1/2}\left|^4 J-1/2_{J} \right\rangle \nonumber \\
&&-\left[\frac{2J+3}{8(J+1)}\right]^{1/2}\left|^4 J+3/2_{J} \right\rangle, \\
\left|D_+;J^{(-1)^{j-1/2}}\right\rangle & = & \left[\frac{2J+3}{8J}\right]^{1/2}\left|^4 J-3/2_{J} \right\rangle 
+\left[\frac{3(2J-1)}{8J}\right]^{1/2}\left|^4 J+1/2_{J} \right\rangle, \\
\left|D_-;J^{(-1)^{j+1/2}}\right\rangle & = & -\left[\frac{3(2J+3)}{8(J+1)}\right]^{1/2}\left|^4 J-1/2_{J} \right\rangle 
-\left[\frac{2J-1}{8(J+1)}\right]^{1/2}\left|^4 J+3/2_{J} \right\rangle .
\end{eqnarray}
The first $q\,g$ states are 
$\left|S_+;\frac{1}{2}^{+} \{2\}\right\rangle$, 
$\left|S_-;\frac{1}{2}^{-} \{2\}\right\rangle$, 
$\left|D_+;\frac{3}{2}^{-} \{3\}\right\rangle$, 
$\left|D_-;\frac{3}{2}^{+} \{3\}\right\rangle$,
$\left|S_+;\frac{3}{2}^{-} \{5\}\right\rangle$, 
$\left|S_-;\frac{3}{2}^{+} \{5\}\right\rangle$.
All other states are characterized by $\left\langle \bm L^2\right\rangle \ge 8$. The parity is reversed for $\bar q\,g$ states. It is not relevant for $\tilde g \, g$ states \cite{zuk83}.

\subsection{States containing only spin-1/2 particles}
\label{tildegtildeg}

The particle-pairs $\tilde g\, \tilde g$, $q_A\, q_A$, $\bar q_A\, \bar q_A$, and  $q_A\, \bar q_A$ are  represented by ordinary states $\left|^{2S+1}L_J\right\rangle$. Let us point out that $\left\langle \bm L^2\right\rangle =0$, 2, 6 for $L=0$, 1, 2, respectively. So, we have considered all states with different values of $J$ and $S$, compatible with $L \le 1$.

Possible combinations of quantum numbers are well-known for pairs containing quarks or antiquarks. For $\tilde g\tilde g$ systems, the parity is given by $P=(-)^{L+1}$ and the charge conjugation by $C=(-)^{L+S}$, as in $q\bar q$ pairs \cite{zuk83}. So the Pauli principle implies that (anti)symmetrical color $\tilde g\tilde g$ states are characterized by $L+S$ even (odd) and $C=+$ ($-$). For $q\bar q$ pairs, there is no such constraint. The examination of all these quantum numbers shows that there is no overlap between the $\left|^{2S+1}L_J\right\rangle$ states for $\tilde g\tilde g$ or $q\bar q$ pairs and the helicity states for $gg$ pairs (given in \cite{heli}). So processes of order 1, or with operators depending on $\bm L^2$, $\bm S^2$ or $\bm J^2$, cannot contribute to transitions between $gg$/$q\bar q$ and $\tilde g\tilde g$ systems. 

\section{Formulas for two different particles} \label{diff}

Various formulas to compute the $T$-matrix and the scattering part of the pressure are given in \cite{LSCB}, but only for two identical particles. We give here the corresponding relations for different masses. For two masses $m_1$ and $m_2$, the asymptotical relative momentum $\bm q$ is related to the center of mass energy $E$ by $E = \epsilon_1(q) + \epsilon_2(q)$ with $\epsilon_i = \sqrt{\bm q\,^2+m_i^2}$, which implies that
\begin{equation}
q(E) = \frac{\sqrt{\left( E^2 - (m_{1} + m_{2})^2 \right) \left( E^2 - (m_{1} - m_{2})^2 \right)}}{2E}.
\label{diff_mass}
\end{equation}
One can check that 
\begin{equation}
\epsilon_1(E) = \frac{E^2+m_1^2-m_2^2}{2E},
\label{eps1}
\end{equation}
with $\epsilon_2(E)$ given by permuting 1 and 2 in the above formula. These quantities are useful to compute the in-medium effects, namely the Bose-enhancement and the Pauli-blocking.

In the results shown below, the main difficulty is to take correctly into account the constraint~(\ref{diff_mass}) into the various integrations on delta-distributions about energy and momentum conservations. As our interaction is spin-blind, we take for the two-body free propagator a spin-independent expression, in the same spirit as in \cite{hugg12}. Using the Blankenbecler-Sugar reduction scheme \cite{blan66}, we find
\begin{equation}
G_0(E;k)=m_1 m_2 \frac{\epsilon_1(k) + \epsilon_2(k)}{2 \epsilon_1(k) \epsilon_2(k)}
\frac{1}{\dfrac{E}{4}-\left(\dfrac{\epsilon_1(k) + \epsilon_2(k)}{2}\right)^2+i \left( \Sigma_1 + \Sigma_2 \right)},
\label{G0}
\end{equation}
in agreement with \cite{mine08} (the normalization is different in this last reference). The parameter $\Sigma_j = \Sigma^R_j + i \Sigma^I_j$ takes into account the gluon self-interaction. As in \cite{LSCB}, we take $\Sigma^I_j=-0.01$~GeV for numerical purposes, and $\Sigma^R_j = 0$ since the real part can be reabsorbed in the effective gluon mass. We have checked that that our results are very similar by using the propagator in the Thompson scheme \cite{thom70}, in agreement with the results of \cite{Mann05}. 

Introducing the notation
\begin{equation}
\Lambda(E)=\frac{E^4-\left( m_1^2-m_2 ^2\right)^2}{E^3},
\label{Lambda}
\end{equation}
the scattering part of (\ref{pot0}) reads, 
\begin{eqnarray}
\Omega_s 
&=& \frac{1}{64 \pi^ 5\beta^ 2} \sum_{\textrm{binary states}} 
\sum_{I} (2 I + 1) \sum_{J^P} (2 J + 1) \sum_{{\cal C}} \text{dim}\, {\cal C} \nonumber \\
&& \left(  \beta \int_{m_1+m_2}^\infty dE\, E^2 q(E) \Lambda(E)  K_1(\beta E)\, {\rm Re} {\cal T}_{\nu}(E; q(E), q(E)) \right.\nonumber \\ \nonumber
&& -\left. \frac{1}{16 \pi^2} \int_{m_1+m_2}^\infty dE \, E^2  q(E)^2  \Lambda(E)^2  K_2(\beta E)\, {\rm Re} {\cal T}_{\nu}(E; q(E), q(E))\, {\rm Im} {\cal T}_{\nu}'(E; q(E), q(E)) \right. \\ 
&& + \left. \frac{1}{16 \pi^2} \int_{m_1+m_2}^\infty dE \, E^2 q(E)^2 \Lambda(E)^2 K_2(\beta E)\, {\rm Re} {\cal T}_{\nu}'(E; q(E), q(E))\, {\rm Im} {\cal T}_{\nu}(E; q(E), q(E))  \right), \nonumber \\
\end{eqnarray}
where the symbol ``prime" is the derivative respective to the energy.
We have checked that this expression is completely equivalent to the one obtained in a Beth-Uhlenbeck formalism, where quantities are written in terms of phase shifts instead of $T$-matrix elements \cite{uhle36}.


\section{Cross section} \label{cross}

Using the formalism of \cite{tayl72}, but adapted for a relativistic kinematics, it can be shown that the differential elastic cross section for a two-body interaction is given by
\begin{equation}
\frac{d\sigma}{d\Omega}(\bm p \leftarrow \bm p_0) = (2\pi)^4 \mu(E)^2 \left| \langle\bm p\, | {\cal T}(E+i0) |\bm p_0 \rangle  \right|^2,
\label{cross1}
\end{equation}
where $\left|\bm p_0\right| = \left|\bm p\,\right|$ and where $\mu(E)=\Lambda(E)/4$ (\ref{Lambda}). The matrix element is evaluated for a given color-isospin channel. As expected, in the nonrelativistic limit, $\mu(E)$ tends to the reduced mass. By integration on the angles, one obtains 
\begin{equation}
\sigma = (2\pi)^5 \mu(E)^2 \int_{-1}^{+1} d(\hat {\bm p}\cdot \hat {\bm p}_0)\left| \langle\bm p\, | {\cal T}(E+i0) |\bm p_0 \rangle  \right|^2.
\label{cross2}
\end{equation}
The ket $|\bm p\, \rangle$ is a plane wave state containing all possible partial components. By decomposing this state into helicity states, we obtain the cross section $\sigma_{J^P}$ for a given color-isospin-$J^P$ channel
\begin{equation}
\sigma_{J^P} = 4\pi^3 \mu(E)^2 \left| {\cal T}_{J^P}(E)\right|^2.
\label{cross3}
\end{equation}
Our purpose is to compare the contributions from various channels to the grand potential at a given temperature for all possible values of the center of mass energy. So, we define a kind of mean cross section $\bar \sigma_{J^P}$ by integrating (\ref{cross3}) on the energy, 
\begin{equation}
\bar \sigma_{J^P} = 4\pi^3 \int_{m_1+m_2}^\infty dE\, \mu(E)^2 \left| {\cal T}_{J^P}(E)\right|^2 
\left[1 \pm f^{1}(\epsilon_1 (E))\right] \left[1 \pm f^2(\epsilon_2 (E))\right].
\label{cross4}
\end{equation}
The in-medium effects are taken into account, depending on the bosonic or fermionic nature of the two interacting particles. Following (\ref{tosolve}) and (\ref{G0}), when $E \to \infty$, we have 
\begin{equation}
{\cal T}_{J^P}(E; q,q'\,) \to V_{J^P}(q,q'\,).
\label{cross5}
\end{equation}
In our model, $V_{J^P}(q,q')$ is essentially the Fourier transform of a Yukawa interaction which behaves like $q^{-2}$. So, the mean cross section, which depends only on $T$, is finite since $\left| {\cal T}_{J^P}(E)\right|^2 \sim E^{-4}$ when $E \gg m_1+m_2$. We have decided to estimate the relative contributions of two channels $J^P$ and $J'^{P'}$ by computing the ratios $\bar \sigma_{J^P}/\bar \sigma_{J'^{P'}}$.

\section{Group-theoretical identities}\label{group}
It is known in group theory that the second order Dynkin indices $I^ R$ in the tensor product $R_i\otimes R_j$  obey a sum rule that can be rewritten as $I^{R_i}\, {\rm dim}\,R_j+I^{R_j}\, {\rm dim}\,R_i=\sum_{{\cal C}} I^{{\cal C}}$, where ${\cal C}$ denotes the representations appearing in the considered tensor product \cite{Ramond}. Using $C_2^{R}=({\dim }\,{adj} /{\dim }\,R) I^R$ \cite{Ramond} an Eq. (\ref{color_scaling}), one straightforwardly shows that
\begin{equation}\label{eq0}
\sum_{\cal C} {\rm dim}{\cal C}\, \kappa_{\cal C}=0.
\end{equation}

If the tensor product involves two identical representations, $R\otimes R$ contains either symmetric (${\cal C}_S$) or antisymmetric (${\cal C}_A$) representations. Then it can be shown that $\sum_{{\cal C}_S} I^{{\cal C}_S}=({\rm dim}\, R+2) I^{R} $ and that $\sum_{{\cal C}_A} I^{{\cal C}_A}=({\rm dim}\, R-2) I^{R} $ \cite{Ramond}. Moreover one can check that
\begin{equation}\label{eqS}
\sum_{{\cal C}_S} {\rm dim}{\cal C}_S\, \kappa_{{\cal C}_S}=\frac{\mathrm{dim}\, R}{2} \frac{C^R_2}{C^{adj}_2},
\end{equation}
\begin{equation}\label{eqA}
\sum_{{\cal C}_A} {\rm dim}{\cal C}_A\, \kappa_{{\cal C}_A}=-\frac{\mathrm{dim}\, R}{2} \frac{C^R_2}{C^{adj}_2}.
\end{equation}

\end{document}